\documentclass[sn-mathphys-ay]{sn-jnl}

\usepackage{geometry}
\usepackage{amsmath,amssymb,amsfonts}
\usepackage{algorithmic}
\usepackage{graphicx}
\usepackage{textcomp}
\usepackage{xcolor}
\usepackage{url}
\usepackage{caption}
\usepackage{subcaption}
\usepackage[most]{tcolorbox}
\usepackage{enumitem,amssymb}
\usepackage{pdflscape}
\usepackage{multirow}
\usepackage{pifont}
\usepackage[normalem]{ulem}
\useunder{\uline}{\ul}{}
\usepackage{array}
\usepackage{hyperref} 
\usepackage{lmodern}
\usepackage{float}

\def\BibTeX{{\rm B\kern-.05em{\sc i\kern-.025em b}\kern-.08em
    T\kern-.1667em\lower.7ex\hbox{E}\kern-.125emX}}

\usepackage{etoolbox}
\usepackage{marginnote}

\newcommand{\filename}[1]{{\texttt{#1}}}
\newcommand{\indicator}[2]{\textit{``r#1: #2''}}
\newcommand{\rqn}[1]{RQ\textsubscript{#1}}
\newcommand{\rp}{\footnote{\url{https://doi.org/10.5281/zenodo.16934566} or \url{https://github.com/biazottoj/rp-tag-debt-bot}}}

\newcommand{\qsWS} {\footnote{\url{https://github.com/biazottoj/tagdebt-bot/wiki/Quick-Start}}} 

\newtcolorbox{findingbox}{
  enhanced,
  colback=gray!10,   
  colframe=black!80,  
  boxrule=0.5pt,     
  arc=3pt,           
  left=8pt, right=8pt, top=6pt, bottom=6pt,
  before skip=6pt, after skip=0pt
}

\begin{document}

\title{TagDebt: A Bot to Support Technical Debt Management}

\author*[a, b]{João Paulo Biazotto}
\email{j.p.biazotto@rug.nl}

\author[a]{Daniel Feitosa}
\email{d.feitosa@rug.nl}

\author[a]{Paris Avgeriou}
\email{p.avgeriou@rug.nl}

\author[b]{Elisa Yumi Nakagawa}
\email{elisa@icmc.usp.br}

\affil[a]{
University of Groningen, The Netherlands}

\affil[b]{
University of São Paulo, Brazil}

\abstract{
\textbf{Context:} Technical debt (TD) is a widely studied metaphor that helps to explain how sub-optimal decisions, which usually have short-term benefits, can harm software maintainability over time. Although incurring TD is not intrinsically bad, tracking and managing TD are crucial to avoid its negative effects. Hence, researchers and practitioners have proposed and developed diverse approaches and tools for managing TD. However, we are still lacking specialized tools for technical debt management (TDM), specifically ones that can be easily integrated into existing development workflows.
\textbf{Objective:}  We present and evaluate TagDebt, a bot that can be integrated within GitHub repositories and automatically assign labels to issues (i.e., SATD or non-SATD). TagDebt helps in the identification of TD (i.e., by looking for self-admitted technical debt (SATD)), leading to more efficient TDM.
\textbf{Methods:} We carried out a Design Science Research study to design and implement TagDebt. For its evaluation, we executed a Technology Acceptance Model (TAM) study through interviews with 16 practitioners, to check the bot's usefulness, ease of use, and contextual factors that might impact the bot's usage (such as team size and practitioners' roles).  
\textbf{Results:} Overall, practitioners found that TagDebt is useful, especially for organizing issues and reducing manual work. Furthermore, they pointed out that the bot is overall easy to use, and its documentation is clear. The analysis also revealed that contextual factors, such as team and codebase size, impact the decision to adopt TagDebt. Finally, several improvements were suggested, such as including features to check and update the source code. 
\textbf{Conclusion:} TagDebt is a proof-of-concept for the development and usage of more specialized tools for TDM. It helps to make TD visible without disrupting existing workflows, which could lead to increased adoption of TDM tools and, consequently, help practitioners avoid the risks of unmanaged TD.
}


\maketitle

\section{Introduction}
\label{sec:intro}

Technical Debt (TD) was first coined by Cunningham~\citep{Cunningham1992} and refers to the future cost of suboptimal decisions made during software development, such as poorly written code (code debt) or inadequate documentation (documentation debt). These decisions may be intentional, made under time-to-market pressure despite known drawbacks, or unintentional, resulting from mistakes or lack of expertise~\citep{Li2015}.

Although TD may not cause immediate harm \citep{Rios2018}, its accumulation over time can severely impact maintainability, increase costs, or even lead to project abandonment~\citep{Guo2011, Avgeriou2023, Avgeriou2025}. To mitigate these risks, technical debt management (TDM) practices have been proposed, covering eight key activities, which involve identification, representation/documentation, communication, monitoring, measurement, prioritization, repayment, and prevention~\citep{Li2015}. In particular, TD \textbf{identification} refers to mapping existing TD items that should be later documented, prioritized, and paid back. In addition, TD \textbf{monitoring} helps to keep TD items visible.

Despite their importance, carrying out TDM activities imposes some challenges~\citep{Avgeriou2021}: they are time-consuming~\citep{Besker2018}, may increase long-term costs~\citep{Alves2016}, and often lack adequate tool support~\citep{Biazotto2025}. Existing tools, plugins, scripts, and bots~\citep{Biazotto2023} are mostly general-purpose solutions (e.g., SonarQube\footnote{\url{https://www.sonarsource.com}}) adapted for TDM rather than being built specifically for it~\citep{Junior2022, Avgeriou2023, Biazotto2025}. In this context, the Manifesto on Reframing Technical Debt~\citep{Avgeriou2025} explicitly calls for specialized tools to make TDM more efficient.

One manifesto recommendation, i.e., ``\textit{to develop workflow-based TDM tools}'', emphasizes the need for solutions that integrate seamlessly into existing development processes. In addition, in a previous study~\citep{Biazotto2025b}, we identified the same concern reported by practitioners, who highlighted that TDM tools must be consistent with their workflows to reduce the friction in tool adoption. However, current tools for TDM (as reported by \cite{Biazotto2023,Silva2022,Junior2022}) usually require updates and changes to existing workflows (e.g., adding the tool as a new step in the CI / CD pipeline). 

The overhead of configuring and adopting such tools, combined with the lack of specific features for TDM, helps explain why the adoption of such tools remains relatively low~\citep{Besker2019, Biazotto2025b}, compromising the uptake of TDM; this characterizes the \textbf{problem} we intend to tackle in this study. Motivated by those gaps, we defined the following research problems: \textit{``(i) Is it possible to develop a specialized TDM tool that is easy to integrate within existing development workflows?''}; and \textit{``(ii) Considering that we can develop such a tool, would practitioners intend to adopt it?''}.

In this context, the main \textbf{objective} of this study is to design, implement, and evaluate a TDM tool that would fit existing workflows and be in line with current practitioner needs. To do so, we carried out design science research (DSR) following the framework proposed by \cite{Wieringa2014}. As a result, in this paper, we introduce TagDebt\footnote{\url{https://github.com/marketplace/tagdebt-bot}}, a bot that provides practitioners with a solution to identify self-admitted technical debt (SATD) items in GitHub issues. To develop and evaluate TagDebt, we followed DSR phases as follows: first, we explored the \textit{social context}, i.e., TDM in practice, to understand challenges, goals, and needs for TDM tools. Then, we explored the \textit{knowledge context} (i.e., research on TDM) to identify existing solutions and designs for the TDM tool. We then \textit{designed} TagDebt, addressing the identified challenges. Finally, we evaluated TagDebt in the \textit{investigation} phase of DSR using the established Technology Acceptance Model (TAM), and discussed practitioners' intention to use TagDebt. 

Beyond proposing another TD detection bot, this paper positions TagDebt as a workflow-integrated TDM tool and investigates how such automation is perceived by practitioners. Our focus is aligned with broader software engineering concerns about the adoption of automation and tools. For instance, we considered prior work showing that perceived usefulness and ease of use are central to technology adoption~\citep{Davis1989} and that automation must be aligned with socio-technical contexts such as team practices and project characteristics~\citep{Elazhary2021}. In this light, TagDebt is designed to (i) embed TD identification into existing GitHub-based workflows with low configuration overhead and few changes to developers’ current practices, and (ii) serve as an object of study on how practitioners perceive automated SATD labeling. Overall, this study provides the following contributions:

\begin{itemize}
    \item \textbf{TagDebt bot}: We introduce TagDebt, a publicly available and fully open-source bot that integrates with GitHub repositories to automatically identify and label issues that report SATD items. TagDebt operationalizes natural language processing (NLP) solutions for SATD identification (in our current implementation, we use the model by \cite{Li2022} and a Large Language Model (LLM)-based approach, using GPT-5-mini). Such solutions (e.g., ML models or LLMs) analyze issue descriptions and add a label to that issue as SATD or non-SATD. Importantly, TagDebt was designed so that the SATD detection component can be easily replaced when desired, enabling the adoption of future models with improved accuracy or models tailored to specific project contexts. However, we note that regardless of the employed detection function, TagDebt will still use natural language as input, and therefore it can only detect SATD, and not TD in other sources like source code. TagDebt is intended to support \textbf{TD identification} and alerting practitioners about the identified TD items. With TagDebt, we embed TDM directly into the development workflow, reducing configuration overhead. Finally, to the best of our knowledge, TagDebt is the first bot that is able to identify TD in GitHub issues without any type of special tag (e.g., \#TODO, \#FIXME). 

    \item \textbf{Empirical evidence on the artifact’s usefulness, ease of use, and adoption contexts}: Through an evaluation with practitioners, we collected empirical evidence about TagDebt. The results highlight the contextual factors that make it more useful, and also highlight contexts that would benefit the most from adopting the bot. Besides, we list improvements that can be implemented in the bot in future work. These contributions help to scope the bot's value to practitioners and improve its transferability to industry.
    
    \item \textbf{An adapted Technology Acceptance Model (TAM) questionnaire for assessing TDM tools}: We applied an adapted version of the TAM method, which contains questions specific to evaluating TDM tools. This methodological contribution can help other researchers and practitioners to replicate our approach to design and validate TDM tools.
\end{itemize}

Such contributions have several \textbf{implications for practitioners}. By integrating into GitHub workflows with minimal configuration effort, TagDebt enables practitioners to detect SATD items earlier (i.e., at the moment issues are created), reducing the risks of debt accumulation. Its configurability allows teams to adapt detection rules to their specific coding standards, documentation practices, and project priorities, thereby improving the precision and relevance of \textbf{TD identification}. Finally, TagDebt minimizes disruption in existing workflows and facilitates adoption in both open-source and industrial contexts.

Regarding \textbf{implications for researchers}, TagDebt offers a reproducible open source platform to study SATD detection in issue tracking. The design of the TAM study presented in this work can be reused to evaluate other TDM tools, supporting comparative studies in the future. In addition, the artifact and its evaluation contribute to the growing body of empirical evidence on the role of automation in TDM, helping shape future research agendas on TD tooling and adoption factors.

The remainder of this paper is organized as follows. Section~\ref{sec:background} presents the background on TD, TDM, and bots in software engineering. Section~\ref{sec:method} shows the research method and describes the phases of the DSR study. Section~\ref{sec:tagdebt} describes TagDebt in detail and Section~\ref{sec:evaluation-design} presents the design of the bot's evaluation. Section~\ref{sec:results} reports the results of the TagDebt evaluation, and such results are discussed in Section~\ref{sec:discussion}. Section~\ref{sec:tov} presents threats to validity and mitigation strategies, and Section~\ref{sec:conclusion} concludes the paper with final remarks and future work.

\section{Related Work}
\label{sec:background}

In this section, we elaborate on the literature related to our study, which encompasses Steps 1 and 2 of our research method (i.e., \textit{Understand Social Context} and \textit{Understand Knowledge Context}). Section~\ref{sec:rw-td-tdm-tools} provides a brief overview of the background of this study, including key aspects of TD, its management, and the use of tools and automation to increase TDM efficiency. In addition, since our study reports on a bot for TDM, in Section~\ref{sec:backg-bots}, we discuss how bots have been explored in TDM and compare them with TagDebt.

\subsection{TD and its Management}
\label{sec:rw-td-tdm-tools}

The TD present in a system can be introduced at various moments of the software development process since debt is not strictly related to the source code. Consequently, there are different TD types based on their source, e.g., source code, architectural decisions, tests, and infrastructure~\citep{Li2015,Alves2016,Junior2022,Bavota2016}. \cite{Li2015} defined nine types of TD:

\begin{itemize}
    \item Requirements TD: It relates to requirements elicitation and can refer to, for instance, a lack of requirements or a misunderstanding of some of them.
    
    \item Architectural TD: bad decisions related to architectural design that could compromise `internal' quality attributes of the software, such as evolvability or maintainability;
    
    \item Design TD: poor decisions made during the design phase, e.g., the division of responsibility among different classes; 
    
    \item Code TD: violations of code quality aspects, for instance, duplicate code, spaghetti code, and duplicated variables; 
    
    \item Test TD: non-optimal decisions taken in elaboration or execution of tests, e.g., lack of tests;
    
    \item Build TD: bad decisions that can harm the software building process, e.g., bad dependency management; 
    
    \item Documentation TD: poor documentation in terms of correctness, completeness, and up-to-date aspects;
    
    \item Infrastructure TD: non-optimal decisions related to the selection of technologies for software development, e.g., using old technologies; and
    
    \item Versioning TD: problems in source code versioning, such as a lack of multi-version support. 
\end{itemize}

To keep TD under control, various activities have been proposed to help practitioners manage it~\citep{McGregor2012, Guo2011, Santos2013, Alves2016}. \cite{Li2015} summarized eight main activities present in the literature. During \textbf{identification}, TD items are detected using several techniques, such as manual inspection or static code analysis. The identified elements can be documented during the \textbf{representation/documentation} activities, and stakeholders are informed about the TD elements during the \textbf{communication} activity. The TD items can then be monitored during the \textbf{monitoring} activity, which ensures that unsolved TD items are under control. The \textbf{measurement} activity is used to quantify the amount of TD in a system and, in turn, enables the \textbf{prioritization} activity, i.e., ranking TD items that must be solved first. TD items can then be fixed during the \textbf{repayment} activity, which also deals with the problems caused by the accumulation of TD. Finally, it is also possible to avoid undesired TD through the \textbf{prevention} activity. 

In addition to the TD types and TD activities described above, it is pertinent to expand on the concept of self-admitted technical debt (SATD). SATD pertains to TD elements that developers themselves formally recognize as such~\citep{Maldonado2015}. For instance, this occurs when developers annotate source code with comments indicating discrepancies or areas in need of changes. SATD refers to a form of documenting TD, which is orthogonal to the underlying TD problem (e.g., a code smell, architectural drift, etc.). In the context of our paper, ``type'' refers to the problem (e.g., code debt, design debt, testing debt, etc.). An SATD item will report a problem, which can be associated with a ``type'' of debt (e.g., code debt). In that sense, SATD is not a type of TD on its own. This is supported by several studies that showed that the nature of problems reported in SATD has also been found in other study contexts, e.g., using static analysis tools, or by interviewing developers. Such a parallel can be observed by comparing the categorization put forward by \cite{Li2015}, in their systematic mapping study, and the categorization put forward by \cite{Bavota2016}, synthesized from looking at SATD items. Both studies report a near-identical top-level categorization. The nature of problems does not change, but how we identify them (through text in comments/issues or using a static source code analysis tool) does.

Although focusing on specific types of TD can bring benefits, a type-agnostic tool can offer broader support for TDM. Reflecting on the challenges of TDM, we considered natural language processing, i.e., SATD detection, as a viable alternative to detect TD items and improve management practices. In this way, as long as a TD item is reported as SATD (i.e., reported in natural language), the bot can identify it and raise awareness of its presence. To achieve this, TagDebt automates the \textbf{identification}, which is typically frequent in TDM processes~\citep{Junior2022,Biazotto2023}. By systematically labeling potential TD items in issues, the tool not only reduces the manual effort usually required for this activity but also provides the necessary information for subsequent management actions, such as prioritization and repayment.

It is worth noting that there is a relationship between our previous work~\citep{Biazotto2025b} and this current study. In our previous work, we conducted a qualitative survey with 103 practitioners (89 valid responses) from industry and OSS. From the answers, we derived a model of practitioners’ concerns and rationales for TDM tool adoption, showing that adoption decisions are deeply related to keeping control over tool execution and outputs. Additionally, we elicited 46 requirements for TDM tools and clustered them into two categories: ``Information to be provided'' and ``Tool usage''. These requirements were intended to guide both the improvement of existing tools and the development of new ones aligned with practitioners' needs. Overall, the study argues that efficient TDM automation depends as much on interaction design and configurability as on detection capability, reinforcing the need for highly customizable, low-disruption, human-in-the-loop tooling.

This current study is a follow-up to the previous one, inspired and guided by its findings. In this current paper, we designed and implemented a tool grounded in this prior knowledge and evaluated it. Therefore, the main contribution of this paper is a novel tool that demonstrates the feasibility of developing a TDM tool aligned with practitioners’ concerns. As a result, we provide new knowledge both by showing the design decisions behind implementing a TDM tool and by presenting practitioners’ opinions about it, which can support the development and improvement of other tools.

\subsection{Bots in SE and TDM}
\label{sec:backg-bots}

The concept of bots can be traced back to 1950~\citep{Turing1950}, and a core aspect related to the development and usage of bots is the effective interaction between humans and machines~\citep{DALE2016,Vinciarelli2015,Zue2000}. Today, advances in artificial intelligence (AI) and natural language processing (NLP) have led to the widespread development of bots across various domains and to new forms of human-machine collaboration to build modern work environments~\citep{Lee2017}.

When it comes to SE, bots help with a variety of tasks, including facilitating communication and helping decision-making processes~\citep{Storey2016}. Bots have become popular on messaging platforms, showing the potential to impact both the social and technical aspects of software development~\citep{Lin2016}. In collaborative development ecosystems, bots automate actions that are typically performed by humans~\citep{Lebeuf2018}, such as repairing failures in GitHub projects~\citep{Urli2018} and coordinating collaborative modeling efforts~\citep{Perez2017}. Similarly, some bots can help automate the deployment and assessment of SE analysis methods~\citep{Beschastnikh2017}. 

When it comes to TDM, there is less support than for other SE tasks. However, there are some bots that can automate tasks related to TD. \cite{Ochoa_2022} proposed BreakBot\footnote{\url{https://github.com/alien-tools/breakbot}}, a bot designed to notify developers if any breaking changes\footnote{A breaking change is a change to supported functionality between released versions of a library that would require a customer to do work in order to upgrade to the newer version. It is advised that library owners (or ``language leads'') document for customers what constitutes approved and supported usage of their libraries.} happen in libraries currently in use. BreakBot does this by performing a static analysis on the library's source code and the source code of the software projects that depend on this library. While BreakBot is concerned with managing build TD (one of the nine types of TD outlined by~\cite{Li2015}), TagDebt has a much broader scope, relying on NLP to identify SATD items reported in issues. This aspect of TagDebt provides greater flexibility to practitioners.

\cite{ref-bot} introduced Refactoring-Bot\footnote{\url{https://github.com/Refactoring-Bot/Refactoring-Bot}}, which analyzes the source code of projects to identify code smells and automatically refactors them. Code smells, a well-known indicator of TD, include suboptimal code, which in turn compromises the design of the software. Thus, the bot supports TDM by addressing code TD and design TD. Refactoring-Bot presents the refactored changes in the form of pull requests, giving developers the chance to review the changes before integrating them into their software. 

\cite{todo-bot} reported on a study examining the impact of TODO Bot\footnote{\url{https://todo.jasonet.co/}} on the projects where it is adopted. TODO Bot creates issues about source code comments that it identifies as containing the ``TODO'' keyword, which indicates postponed activities and things that are not quite right yet and need further work. This is precisely a form of TD, since these comments highlight ``non-optimal or incomplete solutions'' (e.g., \textit{``TODO: I believe the following code is obsolete''}). The findings of this study indicate that the adoption of TODO Bot into software projects encouraged practitioners to introduce TODO comments to increase the visibility of postponed decisions and TD. We note that there is an overlap between the goals of TODO Bot and TagDebt (our bot) in the sense that both can identify TD using natural language. However, while TODO Bot uses source code comments to detect TD, TagDebt is designed to find TD in issues, more specifically in their content (i.e., title, description, or both). Besides, TODO Bot requires developers to use the tag ``TODO'' in each code comment, while TagDebt does not require any specific tag.

\cite{Phaithoon2021} introduced the FixMe bot, which operates by identifying comments in the source code that reference issues that need to be resolved. The bot monitors these referenced issues (including the tag ``FIXME'') and notifies practitioners when the issues are resolved, allowing practitioners to fix/improve the modules depending on the issues. The comments identified by FixMe represent an aspect of TD denoted as ``Hold TD'' to indicate that the code TD is temporarily put ``on hold'' until the related issue is resolved.

Table~\ref{tab:bots-comparison} provides a comparison between existing bots and TagDebt. Specifically in terms of limitations, BreakBot~\citep{Ochoa_2022} is effective for preventing breaking-change problems; however, it targets only build TD and depends on being able to statically analyze both the library and its clients, leaving other TD types and artifacts uncovered. Refactoring-Bot~\citep{ref-bot} operationalizes code-smell removal through automated refactorings, but it is limited to the smells and transformations encoded in the bot and may not generalize to projects in general. As for TODO Bot~\citep{todo-bot}, it helps increase the visibility of self-admitted TD in code comments, but it relies on explicit ``TODO'' markers, meaning that it can miss implicit TD and may create noisy or low-priority issues. FixMe~\citep{Phaithoon2021} similarly depends on explicit tags/references (e.g., FIXME and linked issues) and, while it monitors issue resolution, it does not reason about the TD content reported in issues themselves. TagDebt tackles these limitations by shifting the detection to issues, which are artifacts often used by practitioners to report TD concerns~\citep{KASHIWA2022}. Since TagDebt relies on NLP to classify SATD-related issues, it can capture multiple TD types (as long as they are documented as SATD) and may support early triage via automatic labeling. Compared to the other bots, TagDebt may reduce adoption friction and does not require specific commenting or tagging conventions.

\begin{table}[h]
\footnotesize
\centering
\caption{Comparison of TDM-related bots}
\renewcommand{\arraystretch}{1.05}
\setlength{\tabcolsep}{4pt}
\begin{tabular}{p{1.5cm} p{2cm} p{2cm} p{2.2cm} p{1cm} p{1.5cm} p{2cm}}
\hline
\textbf{Bot} &
\textbf{Goal} &
\textbf{Artifact} &
\textbf{Technique} &
\textbf{TD type(s)} &
\textbf{Output} &
\textbf{Limitation} \\
\hline

BreakBot&
Detect breaking changes and their impact. &
Library PRs and library/client code. &
Static analysis of API changes/usages. &
Build TD. &
PR notifications &
Narrow scope; needs client and library code. \\

Refactoring-Bot &
Find smells and refactor automatically. &
Project source code. &
Smell detection and refactoring rules. &
Code TD, Design TD. &
Opens PRs with refactored code. &
May not fit project constraints. \\

TODO Bot&
Turn TODO comments into issues. &
Code comments with ``TODO''. &
Keyword detection. &
Multiple TD types (issue-reported). &
Creates issues. &
Requires explicit ``TODO'' tag; misses implicit TD and may generate noise. \\

FixMe Bot&
Track on-hold SATD linked to issues. &
Comments (e.g., FIXME) and referenced issues. &
SATD detection and issue monitoring. &
Multiple TD types (issue-reported). &
Alerts when issues close. &
Depends on explicit tags/references; does not analyze issue content itself. \\

TagDebt (this work) &
Label SATD items reported in issues. &
Issue title/description. &
SATD detection in issues. &
Multiple TD types (as long as they are documented as SATD). &
Auto-labels issues. &
Subject to NLP false positives/negatives. \\ 
\hline
\end{tabular}
\label{tab:bots-comparison}
\end{table}

Bots have evolved significantly since their conceptual origin, especially with recent advances in AI, ML, and NLP, which have enabled their widespread adoption across various domains, including SE. In the context of TDM, existing bots such as FixMe work on source code and rely on explicit developer annotations (e.g., \#TODO/\#FIXME) to identify and track TD, primarily supporting the monitoring and repayment of code-level TD. TagDebt, in contrast, is designed to operate on issue trackers and to capture SATD described in natural language (e.g., refactoring requests). Therefore, it complements existing bots by extending TDM automation to earlier stages of development (i.e., issue management). To the best of our knowledge, there are no other TDM bots that classify SATD directly from issue text using an NLP solution, and we explicitly position TagDebt as addressing this gap.

\section{Research Method}
\label{sec:method}
\begin{figure}[!h]
    \centering
    \includegraphics[width=.7\linewidth]{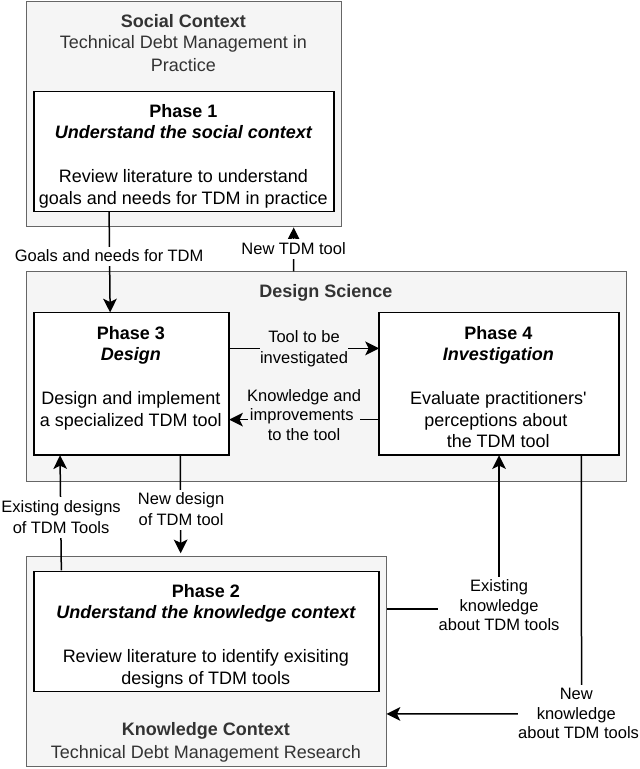}
    \caption{Research Method}
    \label{fig:research-method}
\end{figure}

To guide the design and evaluation of the tool, we adapted \cite{Wieringa2014}'s framework (i.e., DSR) to the specific context of TDM. As Figure~\ref{fig:research-method} shows, this framework helped us establish a structured process that connects the social context (i.e., the needs and demands reported by practitioners) and the knowledge context (i.e., research about TDM). This adaptation ensured that the framework was not applied abstractly, but was operationalized to address the concrete challenges of this study. In summary, we started in the \textbf{``Understand the social context''} phase to identify the challenges and concerns reported by practitioners about TDM. This exploration raised awareness about the problems in TDM. This led to the \textbf{``Understand the knowledge context''} phase to look for existing solutions and designs for the problems. This phase showed how existing tools are used for TDM and their limitations. Next, in the \textbf{``Design''} phase, TagDebt was designed and implemented using the best practices found in the previous phase. Finally, in the \textbf{``Investigation''} phase, TagDebt was evaluated regarding its usefulness, ease of use, contextual factors for its adoption, and potential improvements. Figure~\ref{fig:research-method} summarizes the phases we followed in this study. The boxes represent the phases we performed, while the arrows show the information exchanged between the phases/contexts.

We adopted DSR because our objective is to develop and evaluate an artifact (TagDebt) that enhances support for TDM. Given that TDM involves complex socio-technical challenges, ranging from accurate debt identification to effective prioritization and remediation, there is a clear need for solutions that are both practically viable and theoretically informed. DSR provides a structured framework for iterative design, implementation, and validation of such an artifact, ensuring that it addresses real-world constraints while contributing to the scientific understanding of TDM practices. By focusing on the development of the artifact in terms of both empirical evidence and stakeholder input, our goal is to produce a solution that not only supports current software development workflows but also advances the state of knowledge in the field.

The four phases of our research method are detailed as follows:

\subsection{Understand the Social and Knowledge Contexts}
To explore the \textit{social context}, we used recent literature on TDM. Specifically, we relied on the results of two recent studies we conducted. First, to deepen the understanding of the concerns practitioners have about TDM tooling, we reused the results of a survey with practitioners~\citep{Biazotto2025b}, which yielded 89 valid responses and revealed nine main concerns about TDM tools. The second study was an analysis of practitioner discussions on Stack Exchange~\citep{Biazotto2025} that offered complementary insights into practitioner perceptions. We also reviewed the Manifesto on Reframing Technical Debt~\citep{Avgeriou2025}, which further articulated practitioners' demands, and we examined related work that focuses on tools for TDM. Accordingly, we note that the understanding of the social context did not involve interacting directly with practitioners or conducting additional data collection or analysis. In this phase, we focused only on reviewing existing literature.

To explore the \textit{knowledge context}, we used the results of a systematic mapping study (SMS)~\citep{Biazotto2023} that identified 122 tools that support TDM. This study allowed us to map the current landscape of TDM tools, along with their main characteristics and challenges. Among those, we further investigated tools that were aligned with the main goal of this study (i.e., a tool easily integrated into the software development workflow). As a result, we reviewed the existing bots that can support TDM, including TODO~\citep{todo-bot} and FixMe~\citep{Phaithoon2021}. These sources provided us with the necessary knowledge to design TagDebt. We presented, in Section~\ref{sec:background}, the information we gathered after exploring both social and knowledge contexts.

\subsection{Design}
To design a tool for TDM, we used the requirements described by the practitioners in our previous study~\citep{Biazotto2025b}. We filtered the requirements to build a complete tool use case, as we explain in Section~\ref{sec:tagdebt-requirements}. To implement the bot, we followed the guidelines provided by \cite{Wessel2023}, which support the development of bots for GitHub. That article highlights recurrent problems identified in existing bots, such as their ``noisy'' behavior, where bots leave numerous comments in issues or PRs (pull requests) with excessive information, thereby overwhelming developers and disrupting their workflow. It also addresses the overly humanized comments by bots as well as the complexity and poor documentation for configuring them. That article also suggests seven approaches to avoid common pitfalls, which we used to implement the features of TagDebt.

We approached the implementation of the requirements in an iterative manner, where the implementation of one requirement followed after the completion and manual testing of the previous one. Testing each requirement involved creating test issues to interact with a deployed instance of the bot and to ensure that it works as expected. Section~\ref{sec:tagdebt} describes the design and implementation of TagDebt.

\subsection{Investigation}
\label{sec:sd}
TagDebt aims to provide a specialized solution for TDM. This implies that TagDebt needs to meet the expectations of practitioners. Hence, we assessed to what extent and why practitioners intend to use TagDebt and are willing to adopt it in their daily work. Individual practitioners' \textbf{intention to use} TagDebt can be determined by three factors: \textit{usefulness} of the bot, \textit{ease of use} of the bot, and \textit{contextual factors}. According to \cite{Davis1989}, the usefulness and ease of use play a fundamental role in predicting the degree to which an individual would use a new technology. In addition, the intention to use also depends heavily on contextual factors beyond the influence of a concrete technology, such as the individual's background, environment setup, and facilitating conditions~\citep{Venkatesh2003}. 

The development of TagDebt was mainly guided by a \textit{model of concerns} presented in our previous study~\citep{Biazotto2025b}. A core concern is that practitioners want to keep control over tools' execution and outputs, and this control is mainly related to the interaction between tools and practitioners. To explore how well TagDebt can support this concern, we then decided to focus on evaluating the interaction between TagDebt and practitioners. Similar to previous evaluations~\citep{Manteuffel2016, Kemell2019}, we deemed that a TAM-based evaluation is well-aligned with our main goals (i.e., check the interaction between TagDebt and practitioners and how easily TagDebt is integrated into workflows). The details on how we planned and carried out the evaluation of the TagDebt bot will be further discussed in Section~\ref{sec:evaluation-design}, and Section~\ref{sec:results} reports the results of such evaluation.

\section{TagDebt Bot}
\label{sec:tagdebt}

This section introduces the TagDebt bot, which refers to the Design phase of DSR. First, it is important to explain why we suggested this tool to tackle the lack of specialized TDM tools. As we highlighted in Sections~\ref{sec:intro} and \ref{sec:background}, there are some initiatives to identify TD in issues and pull requests. However, existing ML models are bound to local execution and do not have a well-defined interface to interact with practitioners. Therefore, TagDebt intends to be this interface. In fact, we take advantage of the bot's benefits (e.g., easy integration with development workflows) and combine it with very specialized tools for managing TD (e.g., existing ML models or LLMs for the detection of TD). This goal aligns with the Reframing Technical Debt Manifesto~\citep{Avgeriou2025}, which highlights the need for more specialized tools for TDM. We also identified the same concern in \cite{Biazotto2025b}, which shows the importance of TDM tools that remain consistent with existing workflows. Finally, \cite{Avgeriou2023} stresses that many TD tools remain difficult to adopt because they are not embedded in the workflows practitioners already use, which increases friction and reduces sustained usage. In TagDebt, this concern is addressed by the operationalization of TD support inside the issue-tracking routine in GitHub (rather than disrupting existing workflows). The TagDebt outputs are produced and consumed in the same place where teams already triage, discuss, and organize work (i.e., via native issue artifacts such as labels), ensuring that the identification of TD is delivered as a native workflow action rather than an external report.

\subsection{Requirements Elicitation}
\label{sec:tagdebt-requirements}

To elicit the requirements for TagDebt, we mainly relied on our previous work~\citep{Biazotto2025b}. In that study, we conducted a survey that received 103 responses. The survey was structured in five practical scenarios that illustrate the usage of tools to support TDM. From this study, we collected two main resources:

\begin{itemize}
    \item \textbf{A practical scenario in which a bot labels issues in issue tracking systems:} This scenario was evaluated by the survey respondents as very useful. Since we have the initial evidence of the bot's usefulness, we deemed that implementing this tool could add value for TDM.

    \item \textbf{List of requirements for TDM tools:} The study provides a list of 46 requirements to be implemented in TDM tools. In this paper, we filtered these requirements to select the ones that would compose a complete use case for a GitHub-based bot focused on issue labeling. For instance, in our previous study, we identified that notifying practitioners based on issue labels is a requirement for TDM tools. This requirement complements the automated labeling provided by the bot because it helps developers pay attention to specific issues. Of the 46 requirements presented in the previous study, we selected five for the first version of the bot and imported ipsis literis from the previous study (i.e., no additional data analysis was carried out on the raw data of the previous study).
\end{itemize}

Taking into account the artifacts mentioned previously, five requirements were elicited to develop a full-featured use case for the bot. The requirements are summarized as follows:
\vspace{0.2cm}
\begin{itemize}
    \item \textbf{Requirement 1:} The bot will label issues that contain SATD.
    \item \textbf{Requirement 2:} The bot will send notifications by email.
    \item \textbf{Requirement 3:} The bot will allow customization of which practitioners should be notified.
    \item \textbf{Requirement 4:} The bot will enable customization of the labels that trigger notifications.
    \item \textbf{Requirement 5:} The bot will enable customization of notifications considering the time since an issue was opened (that is, lingering issues).
\end{itemize} 

For \textbf{Requirement 1}, the bot should be able to label issues. This is the core requirement upon which the following four requirements are developed and expanded. It is mainly inspired by the practical scenario presented by \cite{Biazotto2025b}. Furthermore, a core recommendation from \cite{Avgeriou2023} is to strengthen automation for TD identification, so that debt can be found early and with minimal overhead. Therefore, TagDebt implements this recommendation by making automated issue labeling the primary focus of \textbf{Requirement~1}. This decision ensures that identification results are not merely diagnostic signals, but become visible to practitioners, who may use the labels for filtering and triaging TD items. Considering this requirement, the TagDebt bot mainly supports the \textbf{identification} of SATD, since it will process issue titles and descriptions to find SATD items reported in such issues.

In \textbf{Requirement 2}, the bot should send a notification to the practitioners by email; this would help to increase the visibility of issues that contain SATD items. To refine this feature, \textbf{Requirement 3} defines that the bot should allow practitioners to configure who should receive emails about issues. In most software projects, there are usually multiple developers working on different parts of the software. Because of this, the email notifications sent by the bot might only concern a certain group of developers, while others might not be interested in receiving them.

Next, \textbf{Requirement 4} states that the bot should include an option to configure the labels that trigger email notifications. As mentioned earlier in this section, the bot has a SATD detection component (i.e., either the ML model by \cite{Li2022} or GPT-5-mini) that generates and assigns labels to issues based on the title, description, or both. While the current version of the bot provides two types of labels (i.e., SATD or non-SATD), implementing this requirement makes the bot ready to accommodate other SATD detectors that can provide multiple types of labels. By allowing practitioners to select the specific labels that should trigger email notifications, the bot can enable them to prioritize and focus on the issues they find most important. 

To fulfill \textbf{Requirement 5}, the bot should provide an option to configure email notifications considering the time since an issue was opened. Issues that are open and have not been resolved or have not progressed in a long time are called lingering or stale issues. If these lingering issues accumulate, they might diminish the visibility of urgent or higher priority issues. In addition, if these problems persist over time, they can become outdated and confuse developers who join software projects at later stages~\citep{Wessel2023}. Therefore, by developing and implementing this feature in the bot, practitioners can receive periodic updates about lingering issues, enabling them to decide on the appropriate course of action.  

\subsection{Architecture of TagDebt}
\label{sec:tagdebt-architecture}

\begin{figure}[h]
    \centering
    \includegraphics[width=.8\textwidth]{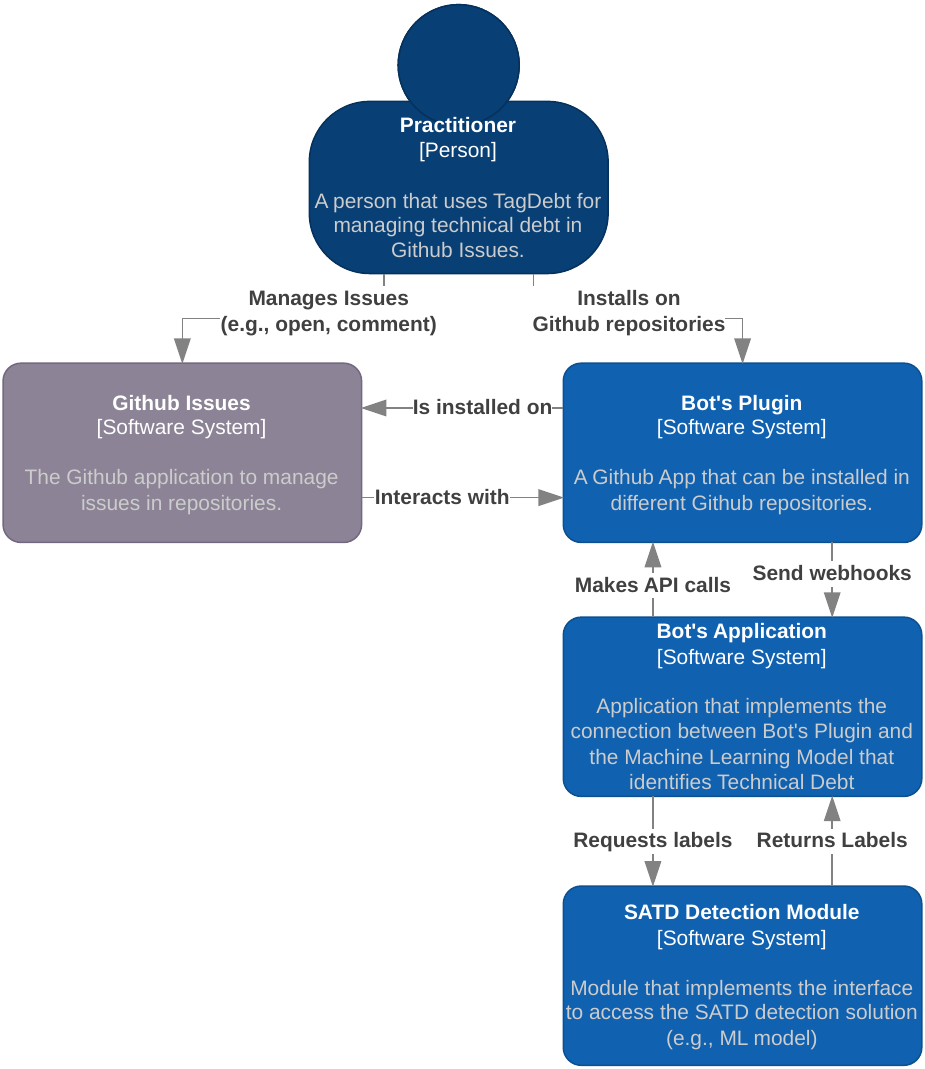}
    \caption{Overview of TagDebt's architecture}
    \label{fig:architecture}
\end{figure}

Figure~\ref{fig:architecture} illustrates the overview of TagDebt's architecture using a Systems Context Diagram from the C4 model\footnote{\url{https://c4model.com/diagrams/system-context}}, highlighting the relationships between the components of the bot and the GitHub repository it interacts with and operates on.

\subsubsection{Bot's Plug-in}
To use TagDebt on a GitHub repository, developers need to install its plug-in. The installation must be carried out by individuals authorized to modify the repository settings, typically the repository maintainers. Once the bot's plug-in is installed, any other project contributor with equal or lower permissions can use and interact with TagDebt.

The bot's plug-in is a GitHub App\footnote{\url{https://developer.github.com/apps/about-apps/}}. When new GitHub issues are created or comments are posted on these issues, the GitHub App triggers webhook notifications (web requests) and forwards them to the bot’s application (its backend). These requests provide all the necessary information in their payload for the bot to perform certain actions.

\begin{figure}[h]
    \centering
    \includegraphics[width=\textwidth]{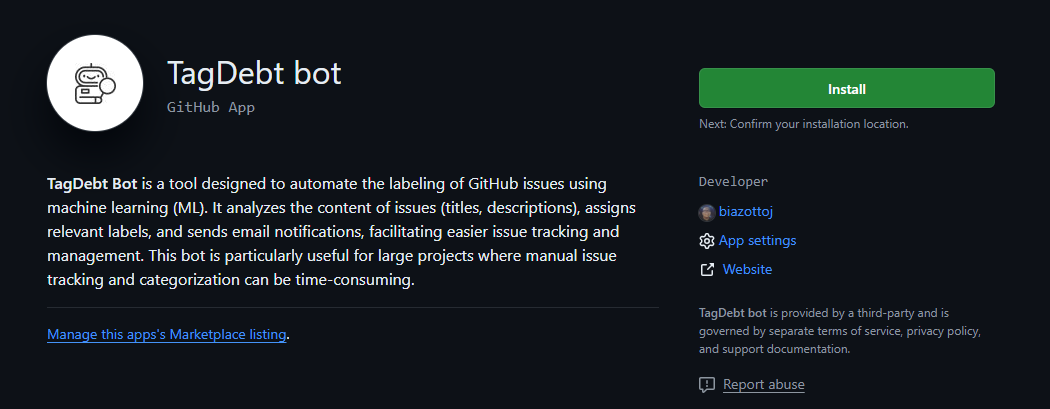}
    \caption{Visualization of TagDebt's GitHub App}
    \label{fig:plugin}
\end{figure}

Besides sending webhook notifications, the bot's plug-in is responsible for establishing the connection between GitHub and the backend application of our bot, acting as an intermediary for any type of communication that happens between the two. The bot's GitHub App is depicted in Figure~\ref{fig:plugin}.

\subsubsection{Bot's Configuration File}
Once the bot’s plug-in is successfully installed on the repository, the bot must be configured. Configuring TagDebt is an essential step, allowing developers to specify how they want the bot to operate in different scenarios. This task can be performed by any project contributor with write access to the repository. During the design of the configuration file, we tried to reduce the complexity as much as possible. This concern is aligned with \cite{Avgeriou2023}, which states that minimizing configuration overhead is crucial to increase the adoption of TDM tools. Specifically, TagDebt is configured through a simple configuration file placed in the repository. Besides, TagDebt provides a default configuration stored in its backend for off-the-shelf use when teams prefer minimal setup. This design keeps setup lightweight and consistent with common development practices (versioned configuration in the repo), reducing the need for specialized tooling expertise.

To configure the bot, a \filename{config.json} file \textbf{must} be included on the main branch of the repository, within a \filename{Bot} folder. However, there is also a default configuration stored in the bot's backend, in case developers want an off-the-shelf solution. We opted for a JSON file format to configure TagDebt because its syntax is straightforward, and JSON files are widely used to configure different kinds of tools and applications~\citep{Harrand2021}. For example, Visual Studio Code (VS Code), one of the most popular code editors developed by Microsoft, uses JSON files named \filename{settings.json} to allow users to customize various settings associated with the editor\footnote{\url{https://code.visualstudio.com/docs/getstarted/settings}}. Thus, it is fair to assume that developers are already familiar with this type of procedure.

\begin{figure}[ht]
    \centering
    \includegraphics[width=0.8\textwidth]{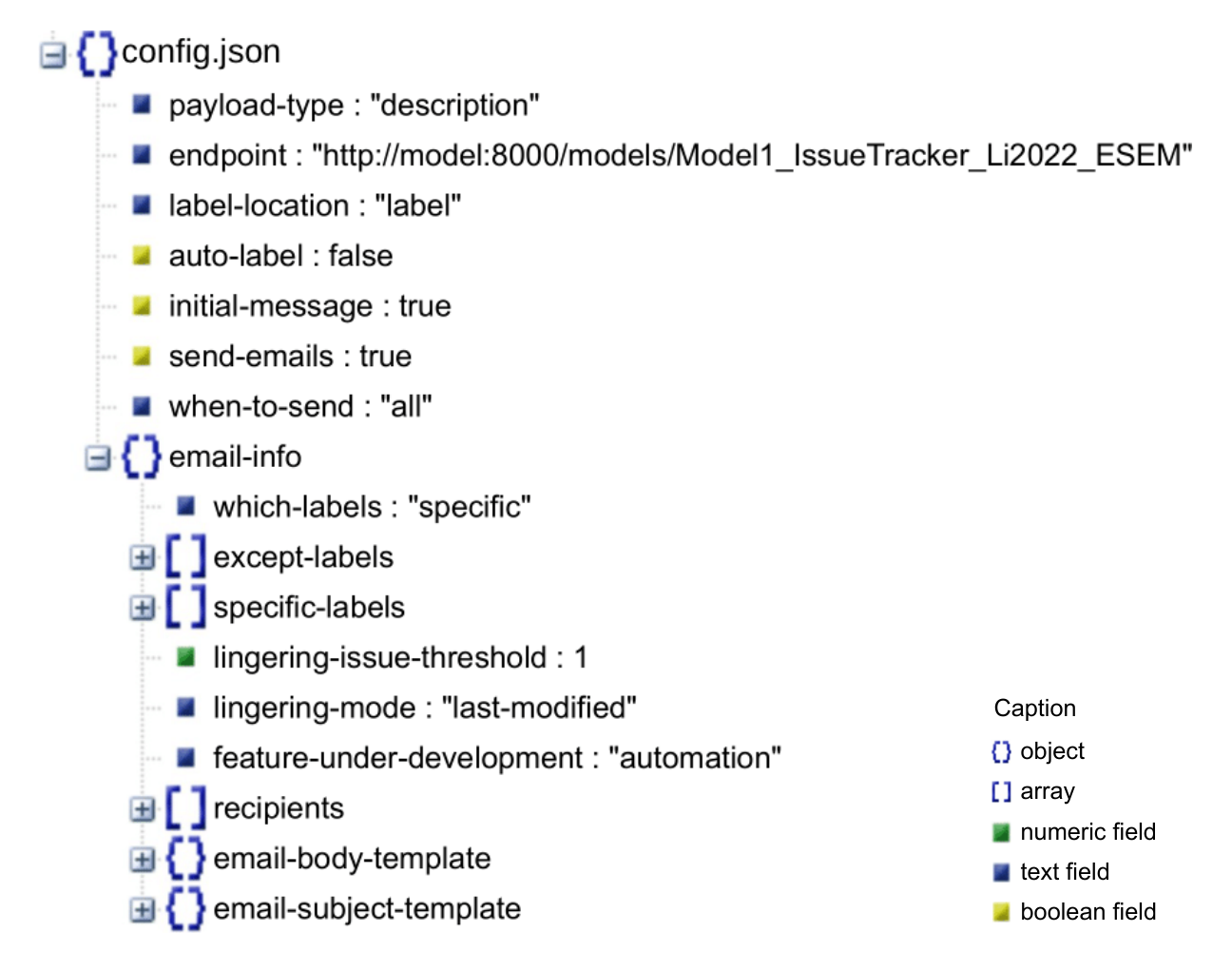}
    \caption{TagDebt's default configuration file}
    \label{fig:tagdebt-config-file}
\end{figure}

The \filename{config.json} file contains multiple configuration options for the bot, each represented by a field paired with its corresponding value. Whenever TagDebt performs an operation, such as processing a command to label an issue using its detection function, it retrieves the latest \filename{config.json} file from the repository. This way, we ensure that the file is always updated. Figure~\ref{fig:tagdebt-config-file} presents the main configuration options within \filename{config.json}. Due to space limitations, some fields are omitted in the figure; however, all options and their descriptions can be found in the bot's Wiki page\footnote{\url{https://github.com/biazottoj/tagdebt-bot/wiki/Documentation}}.

\subsubsection{Bot's Application} 
\begin{figure}
    \centering
    \includegraphics[width=.8\linewidth]{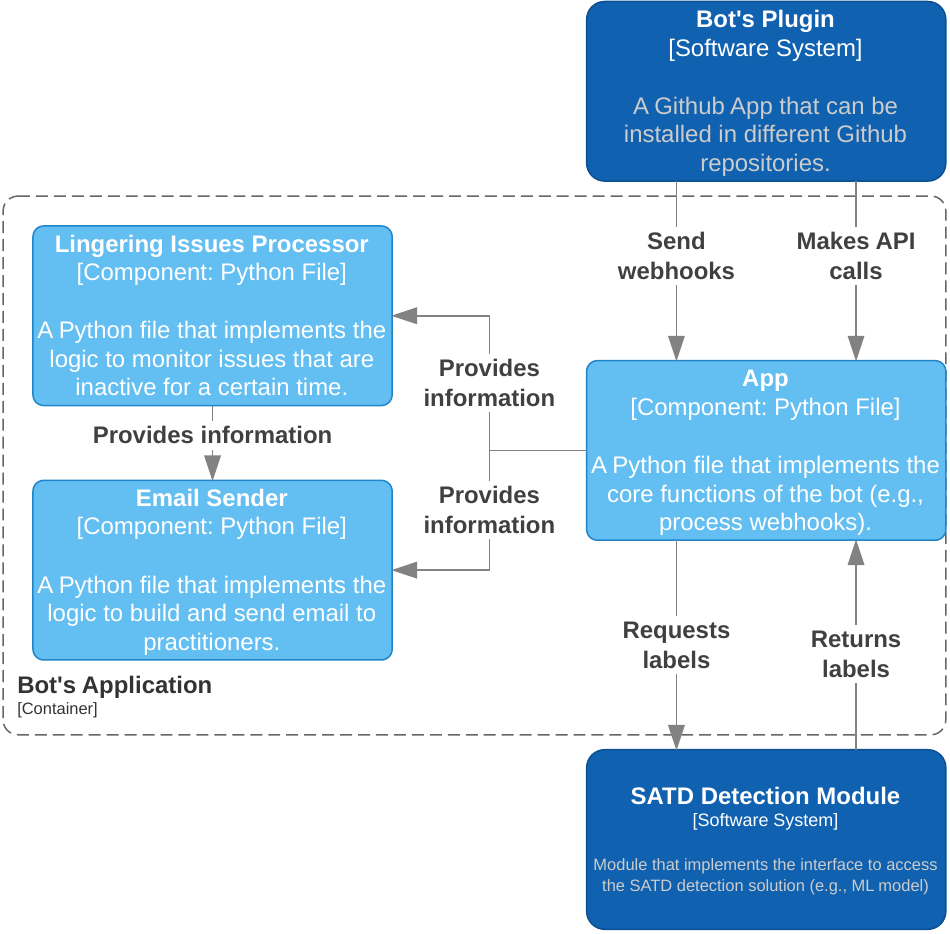}
    \caption{Overview of TagDebt's Application Component}
    \label{fig:bots-app-overview}
\end{figure}

The bot application, or backend, is structured into three different modules, each with a different purpose, to enforce the separation of concerns and enhance modularity. To represent the Bot's application, we used a Component Diagram from the C4 model\footnote{\url{https://c4model.com/diagrams/component}}, as depicted in Figure~\ref{fig:bots-app-overview}. Each module is described below.

\subsubsection*{App}
The \filename{app.py} module coordinates the overall functionality of the bot. It handles the logic for processing incoming GitHub webhooks generated by the bot’s GitHub App. To ensure security, this module validates these requests using an HMAC (Hash-based Message Authentication Code) to confirm they originate from GitHub. Additionally, it retrieves the latest configuration file (\filename{config.json}) stored in the repository, if available, by requesting it from the GitHub API via the bot’s GitHub App. 

Upon receiving a webhook, the module parses the payload to determine the type of event it needs to handle: an issue creation event or an issue comment event, and routes it to the appropriate handler function.
\begin{itemize}
    \item For issue comment events, it examines the content of the comments, and if it identifies a command instructing the bot to label the issue, the module communicates with the SATD Detection Module (i.e., uses either the ML model by \cite{Li2022} or the LLM-based plugin) to classify whether the issue contains TD, and subsequently assigns the predicted label to the issue.
    \item Alternatively, if it identifies a command instructing the bot to label an issue with a specific label, it assigns the provided label directly to the issue, without invoking the detection function.
    \item For issue creation events, the module can post welcome comments, automatically generate labels using the detection function, if configured to do so, and assign them to issues.
\end{itemize}

\subsubsection*{Email Sender}

The component is responsible for sending an email to practitioners. It checks whether the conditions necessary to trigger email notifications are met, based on the \filename{config.json} file, and the information provided by either the main module or the lingering issues processing module. If these conditions are met, it sends the emails accordingly. Such conditions include the labels generated by the detection function and the time since issues were opened or were last active.

\subsubsection*{Lingering Issues Processor}
This component periodically checks for lingering (or stale) issues in the repositories where the bot’s plug-in is installed to notify practitioners about them. This process is scheduled by the main module to run at specific intervals, which can be specified by the practitioner within the source code of the bot’s application before starting it (if not specified, it defaults to running every day). 

\subsubsection{TagDebt's Frontend}
TagDebt is designed to operate autonomously, requiring no human intervention aside from setting it up according to the documentation\qsWS. Therefore, TagDebt does not require a dedicated frontend. Instead, GitHub itself serves as the ``frontend'' for interacting with the bot, and developers send commands to TagDebt through GitHub issue comments.	

It is worth noting that some bots, such as the FixMe GitHub bot~\citep{Phaithoon2021}, employ a frontend to allow developers to configure it through a GUI. While this might represent a more visually appealing way of configuring the bot, it essentially holds the same value as doing it directly in the repository by modifying the \filename{config.json}~\footnote{The complete documentation for the TagDebt bot can be found at \url{https://github.com/biazottoj/tagdebt-bot/wiki/Documentation}} file using GitHub's integrated text editor, as is the case with TagDebt.

\subsection{TagDebt's Execution Flow}
\label{sec:bot-exec-flow}
\begin{figure}[h]
    \centering
    \includegraphics[width=1\textwidth]{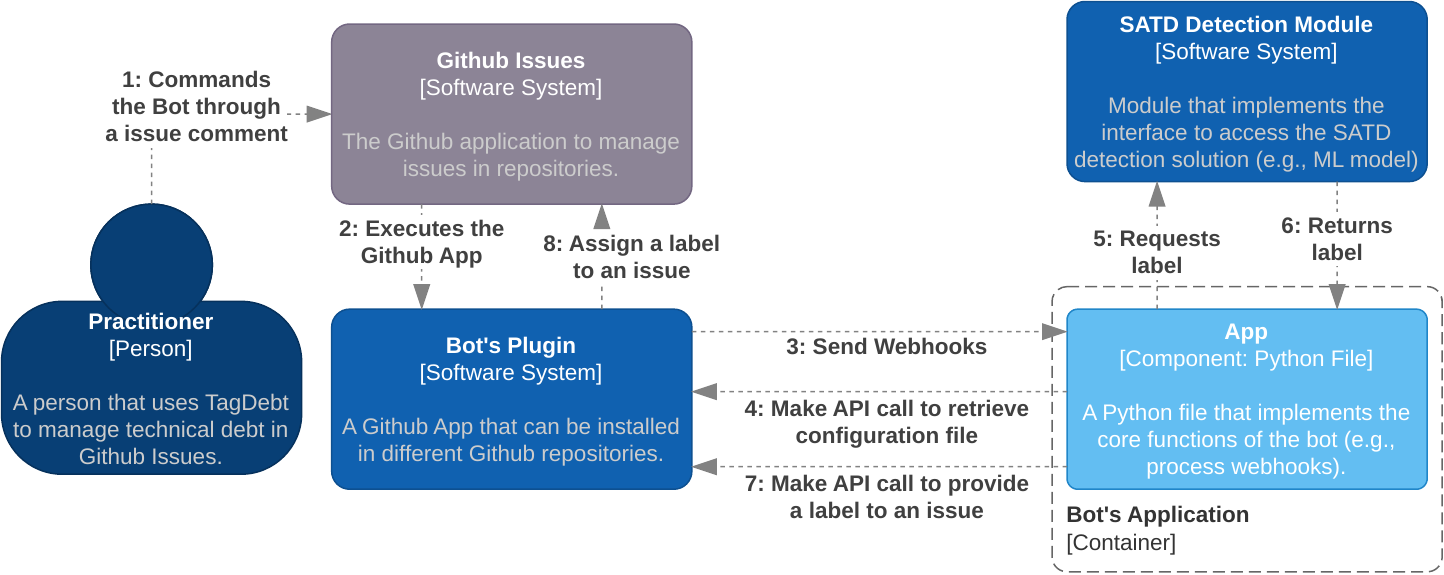}
    \caption{TagDebt's execution flow}
    \label{fig:execution-flow}
\end{figure}

To help understand TagDebt's execution flow and how most of its components interact, we present an example usage scenario, which we will explain using Figure~\ref{fig:execution-flow}. To represent this usage scenario, we used a Dynamic Diagram from the C4 Model\footnote{\url{https://c4model.com/diagrams/dynamic}}. To simplify the understanding of this scenario, we omit the modules related to email sending and the lingering issues checker from the Bot's Application container. The numbered arrows in the figure indicate the sequence of steps performed in this scenario.

Consider a developer who posts a comment on a GitHub issue commanding TagDebt to label that issue (Step 1) (i.e., a \filename{/tdbot label} comment). Once the comment is posted, GitHub Issues executes the bot’s plug-in (Step 2) and sends a webhook notification about this comment to the main App module component (Step 3). The component first validates this webhook event and subsequently requests and retrieves the latest configuration file from the repository by making a request to GitHub’s API via the bot’s plug-in (Step 4).

Next, the App component handles the issue comment event by parsing the content of the comment and identifying the issue labeling command. It then determines which part of the issue (i.e., title, description, or both) should be analyzed to predict whether the issue contains SATD. The part of the issue that should be analyzed is based on the settings specified in the configuration file. Once determined, a request is sent to the SATD Detection Module (step 5). 

In step 6, the SATD Detection Module responds with the predicted label (``\textit{TD}'' or ``\textit{non-TD}''). Following this, the \filename{app.py} module requests GitHub’s API, via the bot’s plug-in, to assign the predicted label to the corresponding issue (step 7). Finally, in Step 8, the Bot's Plugin assigns the provided label to the issue. 

\subsection{Requirements Implementation}
\label{sec:tagdebt-req-implementation}

\begin{figure}
    \centering
    \includegraphics[width=0.7\linewidth]{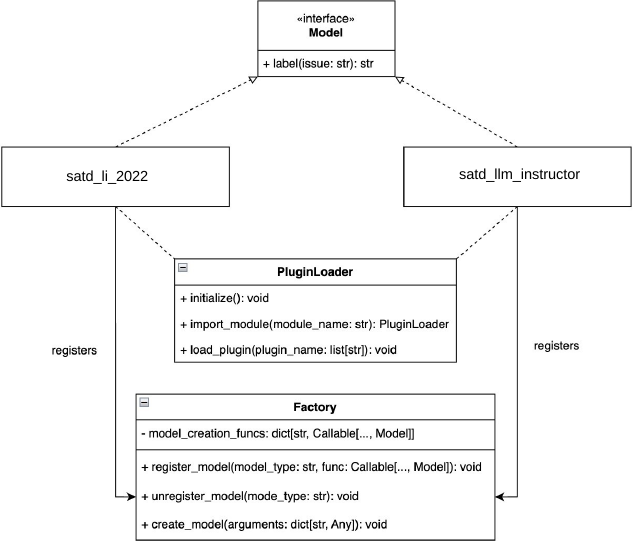}
    \caption{TagDebt's detection module and two examples of compatible NLP solutions (i.e., satd\_li\_2022 and satd\_llm\_instructor)}
    \label{fig:tagdebt-ml-model}
\end{figure}

In this section, we present how we implemented each requirement we described in Section~\ref{sec:tagdebt-requirements}. First, implementing \textbf{Requirement 1} demands deploying or using an NLP solution to identify SATD in issues. We considered that NLP solutions could be deployed and replaced, considering that AI for the identification of TD is constantly evolving~\citep{Li2023b, Avgeriou2025}. Therefore, we decided to implement a factory design pattern~\citep{Gamma1994}, allowing the system to dynamically integrate various NLP solutions without modifying the core application logic. This decision is also aligned with a suggestion from \cite{Avgeriou2023}, which proposed AI-based capabilities as a key enabler for next-generation TDM tooling. In TagDebt, SATD identification is fully based on AI detection functions. Moreover, in TagDebt, the detection function is encapsulated behind a consistent interface (Figure~\ref{fig:tagdebt-ml-model}), allowing practitioners to employ different AI alternatives while preserving the same workflow behavior (i.e., issue labeling).

The current version of TagDebt (v1.1) implements two detection alternatives, i.e., two plugins. The first plugin wraps the Text CNN model proposed by \cite{Li2022}, which was trained to identify SATD in issue tracking systems. In that study, the authors built a dataset of 4,200 issues (23,180 issue sections, including 3,277 SATD sections) and evaluated the optimized Text CNN via stratified 10-fold cross-validation, reporting an average precision of 0.685, recall of 0.689, and F1-score of 0.686. We use \cite{Li2022} as a reasonable default baseline because: (i) it was formally trained to detect SATD in issues, which aligns with the bot’s operating context (issue trackers); (ii) it supports reproducibility, as Li et al. provide a thorough replication package, including examples and training data, which increases the reliability and verifiability of the baseline; and (iii) it offers ease of integration, since its API is simple and reduces engineering overhead to integrate it into a GitHub bot; and (iv) it was a state-of-the-art model at the time we first envisioned the study and implemented the bot (which predated the recent LLM-based advancements in TD detection). At the same time, TagDebt's detection engine is modular and easily replaceable exactly for this reason: different contexts may benefit from different models, and swapping the engine allows practitioners and researchers to adapt TagDebt without being constrained to a single classifier.

The second plugin wraps an LLM-based approach to prompt OpenAI, Anthropic or Google Gemini models. Through this plugin, we also evaluated OpenAI's GPT-5-mini model using the issue dataset from \cite{Li2022}. To this end, we used all SATD entries from \cite{Li2022} (n=1089) and an equally-sized random sample of non-SATD issue descriptions (i.e., N=2178). We needed to sample the non-SATD randomly because there were many more non-SATD samples in the dataset. Specifically, the GPT-5-mini model achieved a precision of 0.759, recall of 0.722, and F1-score of 0.740. The plugin setup is as simple as choosing the model in the configuration file (via a string) and adding the API key to the environment variables. If the developer wishes, the prompt text, API call configurations, and model can be easily adjusted. We chose GPT-5-mini for the evaluation because it is more affordable than the larger counterparts, while providing good performance for SATD classification, e.g., as recently reported by \cite{maarleveld26kubernetes}. Finally, it is important to note that the LLM-based plugin evaluation needs to be interpreted as proof-of-concept regarding potential alternatives that could be used as the detection function of TagDebt. The core contribution of this paper (according to the evaluation presented in Section~\ref{sec:sd}) concerns the mechanisms provided by the bot to integrate SATD detectors into GitHub, as well as its perceived usefulness and ease of use.

Following the C4 model, in Figure~\ref{fig:tagdebt-ml-model}, we present a Code Diagram\footnote{\url{https://c4model.com/diagrams/code}}, using the UML Class Diagram notation. 

The module \filename{factory.py} manages the creation of instances of the detection function (e.g., LLM plugin). The factory has a registry where the plugins are registered with their initialization functions\footnote{We assumed a plugin would have a method/endpoint to be called and trigger a classification, similar to the models implemented by \cite{Li2022} and \cite{Liu2018}.}. The initialization functions are then used to add new detection solutions to the registry, associating them with their respective constructor functions. The detection solutions are instantiated dynamically based on the ``\textit{type}'' field, which defines which detection plugin will be used for the classification, based on the provided configuration file. The bot's backend 
interacts with the detection plugins via a REST API, sending issue data and receiving classification results (i.e., a tag, either ``\textit{TD}'' or ``\textit{non-TD}'').

To implement \textbf{Requirement 2} (i.e., email notifications), we used Python's built-in \filename{smtplib} and the SMTP protocol. This approach allows direct email sending through any compatible server, including Gmail and Outlook. The functionality was encapsulated in a separate module, \filename{emailSender.py}, which defines a function that is called by the main application. Practitioners can configure an SMTP server with credentials securely stored in a hidden file. Each email session establishes and terminates a connection to prevent idle timeouts and unnecessary resource usage. For email content, the MIMEText class from \filename{email.mime.text} is used to create plain text messages. Secure transmission is ensured through TLS encryption, which is widely supported and preferred over SSL due to known vulnerabilities. To configure notifications, two fields can be manipulated in the bot file \filename{config.json}: a) a boolean flag, i.e., \filename{send-emails}, that defines if notifications should be sent; and b) the field \filename{when-to-send}, which specifies when emails should be triggered (e.g., when a certain tag is added to an issue).

As for \textbf{Requirement 3} (i.e., who the bot will notify), the bot’s configuration file contains a field named \filename{recipients} within the \filename{email-info} section. This field contains a list of email addresses of practitioners who wish to receive email notifications from the bot. The decision to allow practitioners to customize their email notification preferences using the existing bot configuration file was based on the aim of incorporating all settings in one place, instead of maintaining a separate file for email addresses, in an effort to reduce the complexity of bots’ configurability highlighted by \cite{Wessel2023}. Practitioners have the flexibility to opt in or out of receiving email notifications from the bot at any time by simply modifying the configuration file in the repository to add or remove their email addresses.

For \textbf{Requirement 4} (i.e., which labels trigger notifications), we use two fields and two sections in the bot’s configuration file (\filename{config.json}):

\begin{itemize}
    \item the \filename{label} field within the \filename{email-body-template} section; and
    
    \item the \filename{label} field within the \filename{email-subject-template} section.
\end{itemize}

To provide further customization, we introduced placeholders (denoted ``/placeholder'') that can be used in the email body templates. The bot email sender parses the templates and replaces the placeholders with data from the labeled issue. For instance, the ``/label'' placeholder is replaced with the label generated by the detection function and assigned by the bot to the issue, and \filename{/issue\_link} is replaced with the hyperlink to the issue where the bot assigned the label.

Finally, to implement \textbf{Requirement 5} (i.e., lingering issues), the bot includes customizable functionality to identify and alert users about issues that remain open or unresolved beyond a defined time threshold. Using the \filename{BackgroundScheduler} class\footnote{\url{https://apscheduler.readthedocs.io/en/3.x/modules/schedulers/background.html}} from the Advanced Python Scheduler library\footnote{\url{https://apscheduler.readthedocs.io}}, the bot runs periodic checks in the background without interfering with other operations. The logic for detecting lingering issues is implemented in the \filename{lingeringIssuesProcessor.py} module, while scheduling is handled in \filename{app.py}, with customizable intervals defined by the \filename{lingering\_check\_frequency} parameter within the \filename{config.json} file. Two key configuration options (i.e., \filename{lingering-issue-threshold} and \filename{lingering-mode}) allow practitioners to specify how many days define an issue as lingering and whether this is based on creation or last modification date. Additionally, the bot supports customizable email templates for both subject and body content, with a dedicated \filename{prepare\_lingering\_email} helper function added to the \filename{emailSender.py} module to format and send these notifications.

\section{Design of TagDebt Bot Evaluation}
\label{sec:evaluation-design}

In this section, we detail the design of TagDebt's evaluation. Specifically, we report its main goal, research questions, and the methods we employed for data collection and analysis.

\subsection{Objective and Research Questions}
\label{sec:sd-objective-rqs}

In this study, our primary goal is to design and develop a tool that is aligned with recent demands for TDM tools, as reported in the literature (e.g., the Manifesto on Reframing Technical Debt~\citep{Avgeriou2025}). Such recommendations involve embedding SATD \textbf{identification} into existing development workflows (in our case, we decided to integrate TagDebt with GitHub issues) with low configuration overhead and without requiring developers to change their current practices. In other words, the central research problem we address is the lack of \emph{workflow-based, adoptable} TDM tools, rather than the lack of high-performing classifiers per se. Besides, most of the design decisions in TagDebt development are focused on practitioners (e.g., do not require changes to developers’ practices). Hence, practitioners’ perceptions of usefulness, ease of use, and contextual factors are the most suitable evaluation criteria to determine whether the artifact actually responds to the identified problems. We structured the \textbf{evaluation objective} according to the Goal-Question-Metric template~\citep{vanSolingen2002} as follows: \textit{``\textbf{Analyze} software practitioners' opinions \textbf{for the purpose of} evaluating TagDebt  \textbf{with respect to} its usefulness, ease of use, and contextual factors \textbf{from the point of view of} practitioners with varied roles \textbf{in the context of} both industrial and open-source software (OSS) development.''}

Based on the goal, the TagDebt bot's evaluation comprised four RQs as follows: 

\vspace{0.3cm}
\noindent
\textbf{\rqn{1} - How do practitioners perceive the usefulness of TagDebt?}
\\Considering the definition of perceived usefulness proposed in \citep{Venkatesh2003}, we investigate to what extent practitioners believe that TagDebt supports TDM and enhances their performance while managing TD.

\vspace{0.3cm}
\noindent
\textbf{\rqn{2} - How do practitioners perceive the ease of use of TagDebt?}
\\Based on the definition of perceived ease of use proposed in \citep{Venkatesh2003}, we want to know to what extent practitioners believe that using TagDebt will not require much effort due to its ease of use.

\vspace{0.3cm}
\noindent
\textbf{\rqn{3} - What are the contextual factors and how do they influence the intention to use TagDebt?}
\\In practice, the usage of a tool often depends on external factors, for example, the nature and size of a software project, the preferred way of working of individuals, or corporate culture and guidelines. This RQ intends to elicit and understand the context that leads to a successful application of the bot in real-world projects. Furthermore, this RQ allows us to evaluate to what extent our results are applicable and transferable to the software industry.

\vspace{0.3cm}
\noindent
\textbf{\rqn{4} - What features could be implemented to improve TagDebt?}
\\This RQ aims to identify concrete suggestions for new features or enhancements that could increase the value of TagDebt. These suggestions may stem from gaps observed by practitioners, unmet needs during usage, or opportunities to better support existing workflows. By understanding how TagDebt can be improved, we can guide its future development, increase user satisfaction, and increase its adoption.

\subsection{Data Collection}
\label{sec:sd-data-collection}

Based on our goal and RQs, we considered multiple acceptance models for evaluating TagDebt, including classical TAM~\citep{Davis1989} and extended models, e.g., UTAUT~\citep{Venkatesh2012}. After analyzing their constructs, we deemed classical TAM sufficient because its core constructs overlap with UTAUT’s core perspectives (i.e., \textit{Perceived Usefulness} and \textit{Perceived Ease of Use} align with \textit{Performance Expectancy} and \textit{Effort Expectancy}), and these constructs are the ones most directly related to our focus on early acceptance and adoption barriers of a GitHub-integrated bot. TAM is widely adopted to understand the acceptance of users of information systems. In SE, TAM has been instrumental in evaluating the adoption of software tools, platforms, and processes by developers. TAM can evaluate two primary factors: usefulness and ease of use. Usefulness reflects the extent to which a user believes that a system enhances job performance, while ease of use refers to the perceived effort required to use the system. According to \cite{Davis1989}, these two factors influence an individual's intention to use a given technology. 

Since this is an exploratory study, we chose interviews instead of a Likert-scale questionnaire (which is more common when applying TAM). This is because a purely quantitative Likert-scale survey would provide limited explanatory depth, which goes against our goal to elicit richer qualitative insights, including contextual factors, concerns, and improvement opportunities that explain why practitioners perceive TagDebt as useful or easy to use (or not).

To apply the TAM-based interviews, we developed a \textbf{questionnaire} based on previous studies (e.g., \citep{Babar2007, Manteuffel2016, Ferreira2024}), adapting the constructs to the context of TDM. Our main objective was to evaluate the extent to which TagDebt is perceived as useful in \textbf{identifying} TD items, reported as SATD, during the software development process. We opted for open-ended questions because they allow us to have in-depth discussions and ask follow-up questions, which help to better understand practitioners' opinions. In Table~\ref{tab:tam-utaut-mapping}, we list the constructs that we used to define each question in our interview guide, as well as their corresponding constructs in UTAUT. For instance, we used the construct ``\textit{\textbf{usefulness}}'' to define \textit{\textbf{U1 - Do you think that the bot would help to identify and monitor TD items more quickly and easily?}}, and then investigated whether TagDebt would help to improve practitioners' productivity.

\begin{table}[h]
\footnotesize
\vspace{-6pt}
\centering
\caption{Mapping between our interview questions and TAM/UTAUT constructs.}
\label{tab:tam-utaut-mapping}
\renewcommand{\arraystretch}{1.05}
\setlength{\tabcolsep}{4pt}
\begin{tabular}{p{0.6cm} p{1.5cm} p{2cm} p{6cm} p{4cm}}
\hline
\textbf{ID} & \textbf{TAM} & \textbf{UTAUT} & \textbf{Rationale} & \textbf{Interview question} \\
\hline

U1 & Perceived Usefulness (PU) & Performance Expectancy (PE) &
Assesses whether embedding TD labeling in GitHub is perceived as reducing the time and effort required to identify TD, which is central to our goal of evaluating perceived usefulness as a proxy for acceptance. &
Do you think that the bot would help to identify and monitor TD items more quickly and easily? \\

U2 & Perceived Usefulness (PU) & Performance Expectancy (PE) &
Captures perceived productivity/coverage gains, supporting our goal of understanding whether practitioners perceive practical value in adopting TagDebt for triage. &
Do you think using the bot would help you to identify and monitor a higher number of TD items? \\

U3 & Perceived Usefulness (PU) & Performance Expectancy (PE) &
Targets perceived effectiveness from the practitioners' perspective (i.e., whether outputs seem to support identifying the \emph{right} items), which informs feasibility and perceived usefulness in workflow contexts. &
Do you think using the bot would help you to correctly identify and monitor the right TD items? \\

U4 & Perceived Usefulness (PU) & Performance Expectancy (PE) &
Provides an overall usefulness judgement to triangulate U1--U3 and summarize perceived value, aligning with our goal of assessing acceptance in an exploratory evaluation. &
Do you think the bot is useful for identifying and monitoring TD items? \\

E1 & Perceived Ease of Use (PEOU) & Effort Expectancy (EE) &
Assesses learnability and onboarding friction, which are critical barriers to adoption for GitHub-based automation and directly relate to our focus on ease of installation and use. &
Was it easy to learn how to operate the bot? \\

E2 & Perceived Ease of Use (PEOU) & Effort Expectancy (EE) &
Configuration overhead is a key adoption barrier for workflow-integrated tools; this question evaluates perceived setup effort and surfaces context constraints affecting feasibility. &
Was it easy to configure the bot? \\

E3 & Perceived Ease of Use (PEOU) & Effort Expectancy (EE) &
Captures memorability and sustained operability (low cognitive overhead), which influences whether a tool remains viable beyond a first trial, consistent with our acceptance-focused goal. &
Do you think it would be easy to remember how to use the bot? \\

E4 & Perceived Ease of Use (PEOU) & Effort Expectancy (EE) &
Documentation quality affects perceived effort and adoption readiness; assessing it supports our goal of identifying concrete improvement points for the bot. &
Do you find the documentation easy to use? \\

Open &
External variables &
Facilitating Conditions &
Elicits missing capabilities, constraints, and improvement opportunities that shape adoption beyond PU/PEOU, consistent with our goal of collecting practitioner concerns and requirements for enhancing TagDebt. &
Is there any other feature you are missing? \\

\hline
\end{tabular}
\end{table}

To evaluate the questionnaire, we performed a pilot study. This study aimed to ensure the understandability, adequacy, and feasibility of our interview instruments prior to conducting full-scale interviews. A practitioner from our personal network was invited to join the pilot study and assigned to use the bot for a few days before the interview. From the pilot study, we observed that the interview would last around 30--40 minutes; we deemed such a duration acceptable to gather relevant data, without tiring the interviewees. In addition, this pilot study allowed us to refine the interview instruments. For example, we initially considered conducting structured interviews, i.e., asking each question individually and in a specified order. Nevertheless, we noted that a semi-structured interview would enable us to collect more insights. We updated the instruments accordingly and used the questions as a checklist (i.e., topics/constructs we should discuss during the interview). The interview transcript from the pilot study was not coded, but it is also available in our replication package.

The \textbf{population} of our study comprises practitioners who are involved in real-world projects, since they are the audience that may potentially use our bot. In terms of recruitment, we adopted a convenience sampling strategy by approaching practitioners from our own professional network and complementing it with a snowballing approach, i.e., asking interviewees to indicate other practitioners who could also contribute to the evaluation. Convenience sampling~\citep{Linåker2015} is a non-probabilistic sampling, in which it is not possible to observe randomness in the selected units from the population. Therefore, there is a trade-off when using convenience sampling: while participants can be recruited quickly, and at low cost and effort, this sampling strategy can limit generalizability~\citep{Baltes2020}.

To mitigate this threat, before recruiting interviewees, we defined a sampling model, according to Table~\ref{tab:demographics}, to ensure diversity in the backgrounds of the respondents. Although we tried to interview as many practitioners as possible, our main goal was to cover a representative range of profiles that could provide varied perspectives on TagDebt. In particular, as a key inclusion criterion, interviewees needed to have industrial software development experience (either current or past); therefore, we did not interview individuals with no industrial experience. Besides, we deemed that two characteristics could directly impact respondents' perceptions about the bot: 
\begin{itemize}
    \item The practitioner's role influences their demands when managing TD~\citep{Rios2020}, and it might impact their perceptions about TagDebt's usefulness. For example, practitioners might focus on information that leads to actionable TDM tasks such as refactoring code, while managers might focus on strategic information that would help with planning and decision-making. To select the roles, we relied on \citep{Rios2020} (reported on page 12, Table~4, in their manuscript), since their work reports a thorough design of a global survey on TD. The selected roles are mainly based on the TD types and cover all the software development life cycle (SDLC). As shown in Table~\ref{tab:demographics}, four roles were considered: Software/System Architect, Test analyst/Q\&A Analyst, Developer/Software Engineer, Project Manager/Team Lead.
    
    \item The practitioner's experience may also influence their perception of tools~\citep{Rios2020}; therefore, we had to balance our sample to consider a wide range of experience levels. To define this aspect, we based the categories on a previous survey we reported in \citep{Biazotto2025b} and defined 3 intervals of experience: \textit{1 to 5 years}, \textit{6-10 years}, and \textit{more than 10 years}. We did not select participants with less than one year of experience because novice practitioners tend to perform simpler tasks that do not involve TDM.
\end{itemize}

\begin{table}[h]
\caption{Sampling Model}
\label{tab:demographics}
\begin{tabular}{
p{4cm}
p{0.1cm}
p{0.1cm}
p{0.1cm}
p{0.1cm}
p{0.1cm}
p{0.1cm}
p{0.1cm}
p{0.1cm}
p{0.1cm}
p{0.2cm}
p{0.2cm}
p{0.2cm}
p{0.2cm}
p{0.2cm}
p{0.15cm}
p{0.5cm}}
\hline
\multicolumn{1}{c}{\multirow{2}{*}{\textbf{Themes}}} & \multicolumn{16}{c}{\textbf{Respondents}} \\
\multicolumn{1}{c}{} & \textbf{r1} & \textbf{r2} & \textbf{r3} & \textbf{r4} & \textbf{r5} & \textbf{r6} & \textbf{r7} & \textbf{r8} & \textbf{r9} & \textbf{r10} & \textbf{r11} & \textbf{r12} & \textbf{r13} & \textbf{r14} & \textbf{r15} & \textbf{r16}\\ \hline
\vspace{0.1cm}
\textbf{Role} &  &  &  &  &  &  &  &  &  &  &  &  &  &  & & \\
Software/System Architect & \checkmark &  &  & &  & & &  &  &  &  &  &  &  & & \\
Test Analyst/Q\&A Analyst &  &  &  &  &  &  &  & \checkmark &  &  &  &  &  &  & & \\
Developer/Software Engineer &  & \checkmark & \checkmark & \checkmark & \checkmark &  &  &  & \checkmark & \checkmark &  & \checkmark & \checkmark & \checkmark & \checkmark & \checkmark\\
Project Manager/Team Lead &  &  &  &  &  & \checkmark & \checkmark &  &  &  &  \checkmark &  &  &  & & \\

\vspace{0.1cm}
\textbf{Experience} &  &  &  &  &  &  &  &  &  &  &  &  &  &  & & \\
1 - 5 years &  &  &  & \checkmark &  & \checkmark & \checkmark & \checkmark &  & \checkmark &  &  &  &  & \checkmark & \\
6 - 10 years & \checkmark & \checkmark & \checkmark &  &  &  &  &  &  \checkmark &  &  &  &  & & & \checkmark\\
more than 10 years &  &  &  &  & \checkmark &  &  &  &  &  &  \checkmark & \checkmark & \checkmark & \checkmark & & \\ \hline

\end{tabular}%

\end{table}

To run the \textbf{interviews}, we asked the practitioners to install and use TagDebt to open (and inspect) issues, without enforcing a specific usage scenario. In particular, we allowed respondents to choose the context that made the most sense to them, such as a real project they work on, a toy repository, or simulated issues created only for the assessment. While this flexibility may reduce comparability across participants, it enabled practitioners to assess the bot in a context that is meaningful and realistic for their own workflow. Moreover, since this is an exploratory study focused on early feedback, we believe that a strictly controlled scenario could have limited the qualitative insights we aimed to collect. Finally, the model provided by \cite{Li2022} was used as the default one for the evaluation. However, we note that practitioners did not evaluate the detection function itself, but the bot's usefulness and ease of use. This means that the detection function does not impact the evaluation setup. 

Given this evaluation setup, we also did not systematically collect detailed information about participants’ current working environments (e.g., organizational maturity, existing TD processes, team distribution, or technology stack). Instead, we deemed it more appropriate to characterize participants in terms of their industrial experience and roles and to rely on their accumulated experience when reflecting on usefulness and ease of use, rather than attempting to map perceptions to a specific organizational setting that may not have been the one used during the evaluation. Nonetheless, we elicited and reported contextual factors that practitioners expect to influence adoption (e.g., team size and project/codebase characteristics), providing a proxy for the contexts in which TagDebt would be more likely to be used.

During the interview, each participant was characterized regarding their role and experience and asked to provide a general impression of using TagDebt. Then, specific questions about the usefulness (RQ1) (e.g., ``\textit{Do you think that the bot would help to identify and monitor TD items more quickly and easily?}'') and ease of use (RQ2) (e.g., ``\textit{Was it easy to learn how to operate the bot?}'') of the bot were asked. Regarding contextual factors (\rqn{3}), they were extracted from follow-up questions to the practitioners. For instance, a respondent made the following statement during the interview: \indicator{3}{I saw that you probably designed it for people who work in large teams. Maybe they have a bottleneck when they have to assign labels.} Hence, we followed up to understand if the team size is a factor that would influence the bot’s usage, which was further confirmed by that respondent while coding the transcript; so, we assigned the code ``\textit{contextual factor - team size}'' to that incident. At the end of the interview, practitioners had the opportunity to suggest improvements for the bot (\rqn{4}). Each interview was recorded with the authorization of the participants, who were informed that their participation was voluntary and they could withdraw their data at any time.

\subsection{Data Analysis}
\label{sec:sd-data-analysis}

To answer our four RQs, we used \textbf{thematic synthesis} to recognize, interpret, and present patterns (themes) within the data. Since the data we collected are mainly qualitative, we deemed this qualitative research synthesis method appropriate for this study. We followed the five steps defined in \citep{Cruzes2011}, as depicted in Figure~\ref{fig:data-analysis} and described below:

\begin{figure}[h]
    \centering
    \includegraphics[width=\linewidth]{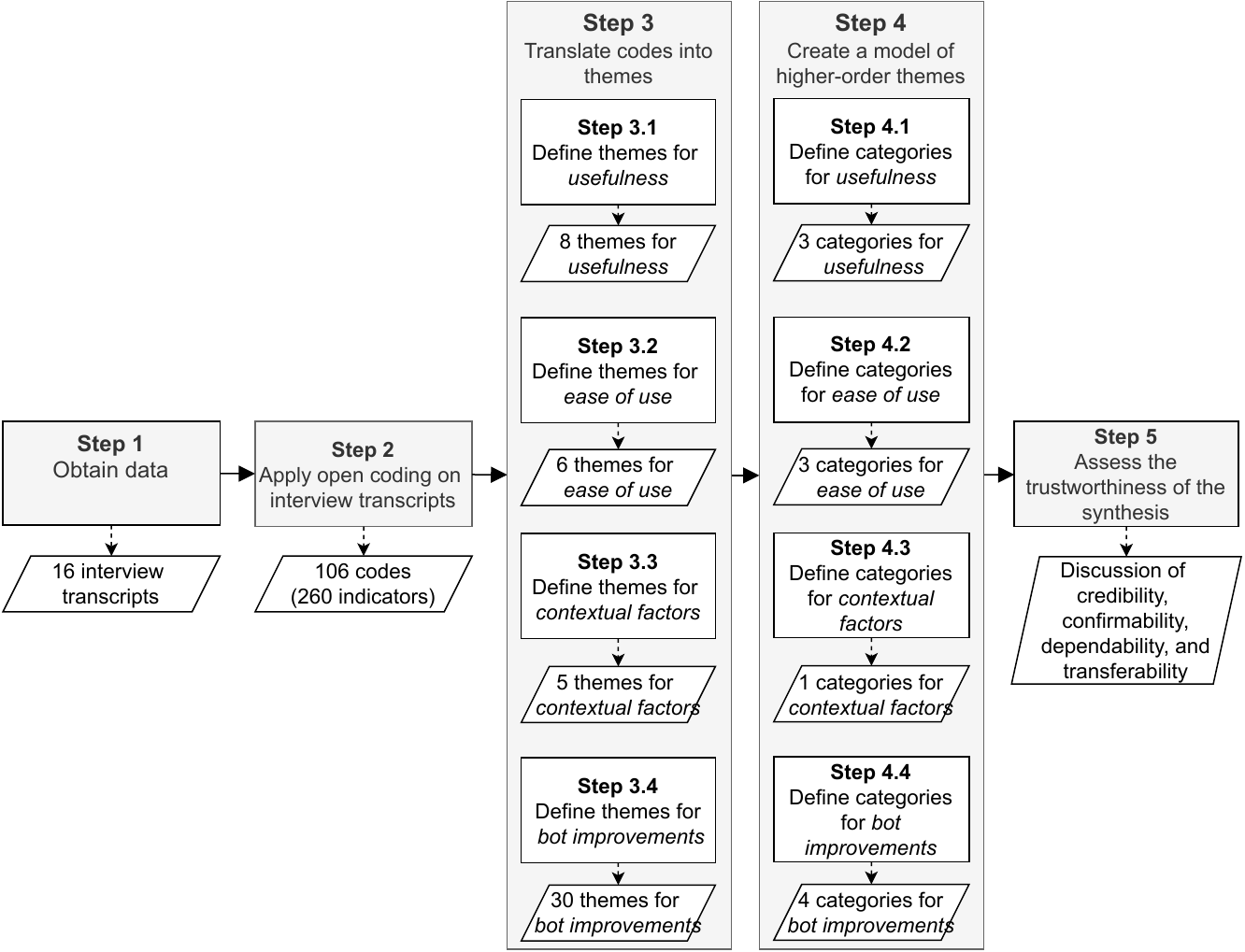}
    \caption{Data analysis method for TagDebt evaluation}
    \label{fig:data-analysis}
\end{figure}

\begin{itemize}

\item \textbf{Step 1 - Obtain data:} After conducting the interviews with the 16 practitioners, we performed the transcription of each interview to make a systematic data extraction possible.

\item \textbf{Step 2 - Apply open coding on interview transcripts:} We applied the open coding technique to break down practitioners' answers 
into smaller pieces of information (referred to as \textit{indicators}). Those were represented by a code (e.g., a word, phrase, label, etc.), which was used to navigate through different indicators and compare them using constant comparison. For example, one practitioner stated that: \indicator{1}{And then eventually you can assign based on the label to different people. And based on that, you can prioritize and organize your priorities.} Hence, we assigned the code \textit{usefulness - the bot could support prioritization} to that indicator, which helped us compare that indicator with others. 

To code the data, we used an integrated approach~\citep{Cruzes2011}. Initially, two authors coded a sample of 16 answers (which were randomly selected), generating as many codes as possible from the data. After that, they discussed the generated codes until reaching an agreement. Next, the first author coded all the transcripts, using both the previously generated set of codes and proposing new codes, as necessary to better represent the indicators. After that, the four authors (the two authors who were involved in the open coding, plus the last two authors) discussed the codes and reached an agreement on the indicators and codes to represent them. Considering that we had three aspects to consider (i.e., usefulness, ease of use, and contextual factors), we assigned an aspect to the codes (e.g., ``\textit{usefulness - team size}'') to help us cluster the information about those aspects. The application of open coding resulted in a set of 106 codes and 260 indicators. Table~\ref{tab:open-coding-example} presents a few examples of these codes and indicators.

\begin{table}[!h]
\centering
\caption{Example of coding process}
\footnotesize
\begin{tabular}{p{3.0cm}p{2cm}p{2cm}p{2cm}p{2cm}}
\label{tab:open-coding-example}
\\ \hline
\textbf{answer} & \textbf{author 1} & \textbf{author 2} & \textbf{final code} \\ \hline
OK, first of all, classification and categorization, but that’s general for any kind of labels or tags. The thing is you can easily group them and see the status of your project, the status of your code. & simplifies aggregation & helps in classifying issues & usefulness - helps in classifying issues.  \\
Yes, it was. It was very easy, and I could understand from the first reading what you wanted with each option there. & good documentation & documentation is simple & usefulness - documentation is simple   \\\hline

\end{tabular}

\end{table}

\item \textbf{Step 3 - Translate codes into themes:} A theme emerges through the process of coding, categorization, and reflective analysis, and represents an abstract construct that imbues recurring experiences with meaning and coherence. In this step, we mainly used selective and theoretical coding, which enabled us to translate codes into themes. The emphasis of this approach is constant comparison, which helps to mitigate potential biases and enhance data exploration. For example, after performing open coding, we obtained codes such as ``\textit{usefulness - bot can support planning TD repayment}'' and ``\textit{usefulness - the bot can support with issue prioritization}.'' During selective coding, these codes were grouped, and the theme \textit{``Prioritizing TD items''} was defined. In theoretical coding, we then grouped themes into broader categories. For instance, the themes \textit{``Prioritize TD items''} and \textit{``Increase TD awareness''} were grouped under the category \textit{``TD Visibility''}. We applied this same procedure to all RQs: for each RQ, we independently translated its codes into themes and then into categories (i.e., Steps 3.1 to 3.4 in Figure~\ref{fig:data-analysis}) so that the themes emerged individually and specifically for each RQ. It is important to note that we did not adhere strictly to all steps of Grounded Theory~\citep{Strauss1990}, and therefore, the resulting themes cannot be claimed as a fully developed theory. Interactive data collection and analysis were not employed; thus, theoretical saturation cannot be guaranteed, a point further discussed in Section~\ref{sec:tov}.

\item \textbf{Step 4 - Create a model of higher-order themes:} The themes that emerged in the previous step were explored and interpreted to create a model consisting of higher-order themes. In this study, we propose a set of categories of themes related to the three aspects of adopting TagDebt (usefulness, ease of use, and contextual factors). For instance, themes such as \textit{``Reduces repetitive or manual work''} and \textit{``Helps prevent human errors''} were grouped under the category \textit{``Developer Support''}. We defined the categories individually for each one of the RQs (i.e., Steps 4.1 to 4.4 in Figure~\ref{fig:data-analysis}). Section~\ref{sec:results} presents and discusses these categories. 

\item \textbf{Step 5 - Assess the trustworthiness of the synthesis:} Assessing the trustworthiness (or reliability) of interpretations derived from thematic synthesis involves constructing arguments supporting the most plausible interpretations. Research findings do not have a singular correct interpretation but rather the most probable interpretation from a specific standpoint. Therefore, enhancing trustworthiness entails presenting findings in a manner that encourages readers to consider alternative interpretations. To evaluate the trustworthiness of the results, we examined the concepts of \textit{credibility}, \textit{confirmability}, \textit{dependability}, and \textit{transferability}. Section~\ref{sec:tov} details this evaluation.

\end{itemize}

\section{Results of TagDebt Evaluation}
\label{sec:results}

\begin{figure}[h]
    \centering
    \includegraphics[width=1.2\textwidth]{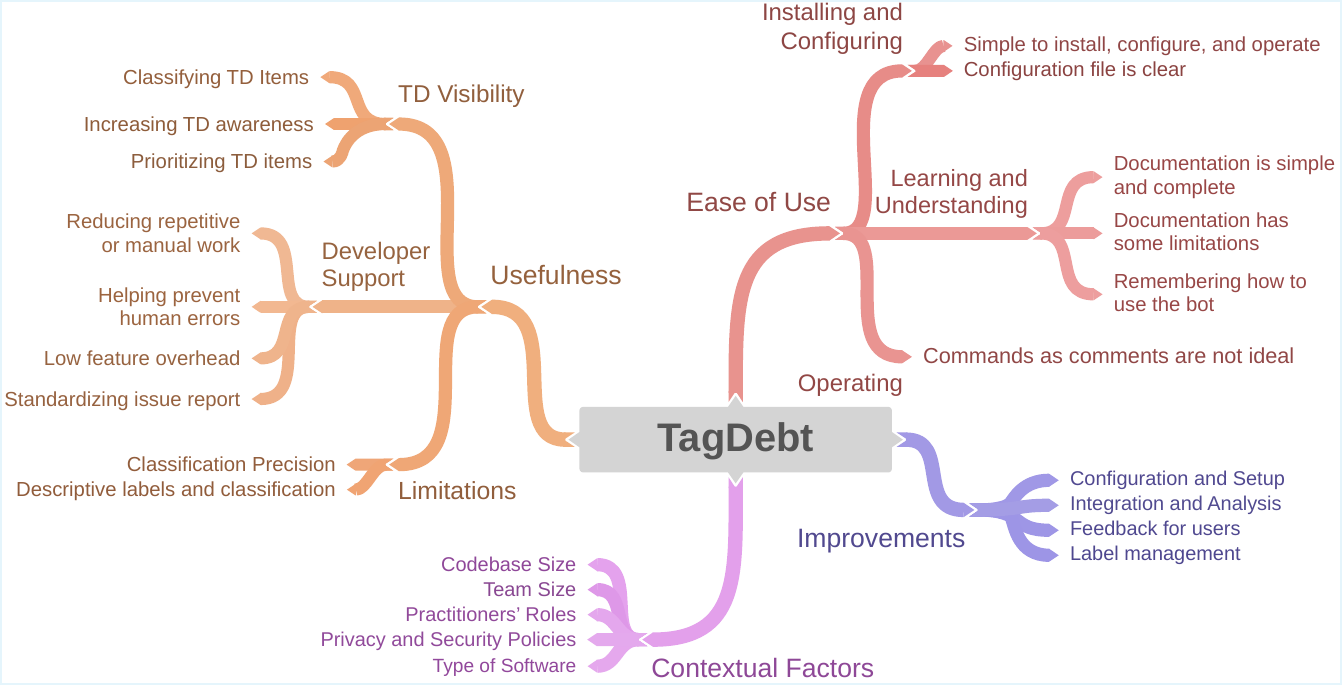}
    \caption{Summary of the themes identified for usefulness (\rqn{1}), ease of use (\rqn{2}), contextual factors (\rqn{3}), and improvements (\rqn{4})}
    \label{fig:results-summary}
\end{figure}

This section presents the results for the \textbf{Investigation} phase of DSR. Overall, the results are based on the themes related to usefulness, ease of use, contextual factors, and improvements. Figure~\ref{fig:results-summary} summarizes the themes identified for the four RQs, which are further discussed in Sections~\ref{sec:results-usefulness} to \ref{sec:results-improvements}.

\subsection{Usefulness of TagDebt}
\label{sec:results-usefulness}
To answer \rqn{1}, we investigated how practitioners perceive the usefulness of the TagDebt bot. Overall, participants identified multiple ways in which TagDebt adds value to TDM. The topics related to usefulness were classified into three categories: (i) \textit{TD Visibility}; (ii) \textit{Developer Support}; and (iii) \textit{Limitations}. Table~\ref{tab:themes-usefulness} shows the themes within each category.

\begin{table}[h]
\caption{Themes related to usefulness, the respondents supporting each theme, and the total of indicators per theme}
\label{tab:themes-usefulness}
\begin{tabular}{
p{4cm}
p{0.1cm}
p{0.1cm}
p{0.1cm}
p{0.1cm}
p{0.1cm}
p{0.1cm}
p{0.1cm}
p{0.1cm}
p{0.1cm}
p{0.2cm}
p{0.2cm}
p{0.2cm}
p{0.2cm}
p{0.215cm}
p{0.215cm}
p{0.215cm}
p{0.15cm}}
\hline
\multicolumn{1}{c}{\multirow{2}{*}{\textbf{Themes}}} & \multicolumn{16}{c}{\textbf{Respondents}} \\
\multicolumn{1}{c}{} & \textbf{r1} & \textbf{r2} & \textbf{r3} & \textbf{r4} & \textbf{r5} & \textbf{r6} & \textbf{r7} & \textbf{r8} & \textbf{r9} & \textbf{r10} & \textbf{r11} & \textbf{r12} & \textbf{r13} & \textbf{r14} & \textbf{r15} & \textbf{r16} & \textbf{Ind.} \\ \hline
\vspace{0.1cm}
\textbf{TD Visibility} &  &  &  &  &  &  &  &  &  &  &  &  &  &  &  &  & \\
Classifying TD Items & \checkmark & \checkmark &  & \checkmark & \checkmark & \checkmark & \checkmark & \checkmark &  & \checkmark & \checkmark & \checkmark & \checkmark &  & \checkmark & \checkmark & 25 \\
Increasing TD awareness &  &  &  &  &  & \checkmark & \checkmark & \checkmark & \checkmark &  & \checkmark &  & \checkmark & \checkmark & \checkmark & \checkmark & 14 \\
Prioritizing TD items & \checkmark & \checkmark &  & \checkmark & \checkmark & \checkmark &  &  &  & \checkmark & \checkmark &  & \checkmark & \checkmark & & & 12 \\

\vspace{0.1cm}
\textbf{Developer Support} &  &  &  &  &  &  &  &  &  &  &  &  &  &  &  &  \\
Reducing repetitive or manual work & \checkmark & \checkmark & \checkmark & \checkmark & \checkmark &  & \checkmark & \checkmark & \checkmark & \checkmark & \checkmark & \checkmark &  &  &  & \checkmark & 25 \\
Helping prevent human errors &  &  &  &  & \checkmark &  & \checkmark & \checkmark &  & \checkmark & \checkmark &  &  &  &  & \checkmark & 7 \\
Standardizing issue reports & \checkmark &  &  &  &  &  &  &  &  & \checkmark & \checkmark &  &  & \checkmark &  & \checkmark & 7 \\
Low feature overhead &  &  & \checkmark &  &  &  & \checkmark & \checkmark &  & \checkmark &  &  &  &  & & & 4 \\

\vspace{0.1cm}
\textbf{Limitations} &  &  &  &  &  &  &  &  &  &  &  &  &  &  &  &  \\
Classification Precision & \checkmark & \checkmark &  & \checkmark & \checkmark &  &  & \checkmark &  & \checkmark & \checkmark & \checkmark & \checkmark & & & \checkmark & 14 \\

Descriptive labels and classification &  & \checkmark & \checkmark &  & \checkmark & \checkmark & \checkmark &  &  &  &  &  & \checkmark &  & &  & 8 \\\hline

\end{tabular}%

\end{table}

Three themes were identified in the category \textbf{\underline{TD Visibility}}. \textbf{Classifying TD items} (supported by 13/16 respondents; 25 indicators) relates to the main feature of the bot: \indicator{1}{[It helps in] classification, categorization… that’s general for any kind of labels or tags}. In addition to classification, the bot helps to keep TD visible over time, since labeled TD items can be aggregated, helping developers filter the issues: \indicator{5}{in just one click [using GitHub's features] you can have all the lists [of issues per label] otherwise you would have to go issue-by-issue to understand the nature of each [TD] item}. Finally, participants also reported that the bot could improve the visibility of TD, helping to pinpoint maintenance hotspots: \indicator{10}{[I can] see where there are more pending tasks or which modules require more attention}. 

TagDebt also helps practitioners to avoid the possible drawbacks of TD accumulation: \indicator{8}{You end up not considering [TD], and it becomes a snowball. You don’t understand the impact… I think having a tool to alert you or help you manage all this is really helpful.} The lack of visibility often leads to TD being forgotten, and TagDebt helps developers to stay aware of TD items that will be managed at a more convenient time: \indicator{9}{After the technical debt is identified, it may get lost… the team may just be focused on delivering, delivering, delivering… and after a couple of months, we won’t even remember it anymore.} 

TagDebt also helps in \textbf{Increasing TD awareness} (supported by 9/16 respondents; 14 indicators), since TD items become visible to developers (through labeled issues). Such awareness helps practitioners to make more informed decisions during the development process: \indicator{6}{[It could] help with making us more aware to maybe reconsider the issue or look into it and have a quick discussion about [whether] this can introduce technical debt.} Some participants also mentioned that the awareness promoted by TagDebt can help reduce the accumulation of TD over time: \indicator{6}{I would say it would not help in the shorter term, but maybe for the longer term because it helps you introduce less technical debt.} Furthermore, the awareness of TD items helps teams better plan maintenance actions earlier in the development process: \indicator{6}{[The bot would be useful] even before the design, helping developers avoid or reduce TD early on.}

For supporting \textbf{Prioritizing TD Items} (supported by 9/16 respondents; 12 indicators), TagDebt helps group the TD items and helps practitioners to identify critical issues that need to be addressed first: \indicator{2}{once [the bot] understands what technical debt is, maybe we can identify what is critical, what is minor.} Finally, by easily analyzing the issues labeled by TagDebt, TD items can be assigned to the right stakeholder, who will decide how to proceed depending on their priorities: \indicator{1}{you can assign [issues] to different people, and based on that, you can prioritize and organize... Your priorities.}

Regarding \textbf{\underline{Developer Support}}, TagDebt can automate daily tasks and reduce the cognitive load of developers. Since TagDebt automatically labels issues, it helps in \textbf{Reducing repetitive or manual work} (supported by 12/16 respondents; 25 indicators), such as manually identifying TD (which is time-consuming): \indicator{9}{I believe it is a very useful tool thinking about this technical debt context, because it ends up being a very manual process and for that reason very flawed}. Once TagDebt reduces the time to carry out simpler tasks (that is, assign labels), developers can focus on tasks that add more value to companies: \indicator{8}{all this manual work often takes time and doesn’t require specific knowledge… a tool that can remove these manual steps and leave decision-making to the developer would optimize the work a lot.} Therefore, TagDebt helps practitioners increase their productivity and reduce the time involved in TDM: \indicator{12}{If we could, for example, pass the labels and the description of what each label is, like which problems they relate to, I think it would save a lot of time… it would help in mapping these things… I see it solving an agility problem in the team’s daily routine}.

TagDebt is also useful in \textbf{Helping prevent human errors} (supported by 6/16 respondents; 7 indicators), since it can be used as a double-check mechanism for manually-assigned labels: \indicator{5}{The bot could be used as a double check… the final answer will be the developer’s one.} The bot can trigger developers to reflect on their own labels and improve developers' precision in identifying TD: \indicator{7}{If you classify the message with label A but the bot labels with label B, it will force you to think critically… oh, why did the bot classify with label B? Maybe I have to look again and avoid the error.} Participants, nonetheless, note that TagDebt would be more useful as a supporting tool, rather than replacing human judgment. Therefore, it can add more value in situations where developers may overlook tagging an issue: \indicator{5}{In some cases, the developer might not tag the issue at all because he is not paying attention, he forgets… in those cases, the bot could help. It is helpful.} An interviewee also described that the content of issues might be ambiguous, so TagDebt could perform a first check: \indicator{10}{It happens a lot [...] you have a bunch of open debt… sometimes the text is ambiguous. I think if there were a tool that could facilitate this or do it automatically, it would definitely make life easier for everyone.}

\textbf{Standardizing issue reports} (supported by 5/16 respondents; 7 indicators) is another benefit that practitioners identified when using TagDebt: \indicator{1}{If you are a software house, you are a vendor, you need to have a standardized template, and you cannot permit your developers to open an issue with just text.} Practitioners imply that the bot can help teams organically improve issue descriptions, and this can help them to define standard templates for issue reports: \indicator{10}{I see it [the bot] as a template, like for opening PRs. In all companies I have worked for, there is always a PR template - it is how we agree to describe what is being done in that pull request. It’s a great way to standardize things, you know?}

One of TagDebt's characteristics that was also highlighted is the \textbf{Low feature overhead} (supported by 4/16 respondents; 4 indicators), since it avoids unnecessary complexity or excessive features: \indicator{7}{ I think it is great in their specific use case. Labeling the issue...} Such specific functionality is a strength, since it avoids excessive information load that a huge set of metrics and diagnoses could cause: \indicator{8}{It's not really for diagnostic suggestions. It’s something very specific. I think that is OK}.

There were also several critical reactions to the TagDebt bot. Such criticism was mostly related to \textbf{Classification Precision} (supported by 10/16 respondents; 14 indicators) and \textbf{Descriptive labels and classification}. The first theme is that the bot's precision is a concern, and it may reduce its usefulness in certain contexts: \indicator{1}{I would not rely on the auto label because otherwise, I would probably spend most of the time changing labels because I meant something.} This concern with precision reduces the trust in the bot, and is taken into account when using TagDebt: \indicator{2}{could be a good starting point, but it first needs to 'gain the team’s trust'}. We clustered such themes into the category \textbf{Limitations}, since we deemed it key that practitioners and researchers also learn from the drawbacks present in our bot.

More \textbf{descriptive labels and classification} (supported by 6/16 respondents; 8 indicators) mainly calls for more explainability about how the bot classifies an issue. First, the bot could provide a more descriptive label, explaining why the issue has a TD item: \indicator{5}{They just say something like ‘td’ or ‘non-TD’, but they don’t explain what [this means]}. The lack of such explanations may lead to the labels reducing the value added by the bot: \indicator{6}{If there were explanations, then that would give much more direction. Because now... It’s up to the developer to figure out where the technical debt is introduced or why it got that label.} Finally, understanding the reasoning behind the classification was mentioned as relevant: \indicator{7}{See why it made that classification? That would be great.} This shows that more informative tags could help clarify the bot’s decisions and improve its usefulness.

\begin{findingbox}
\textbf{Answer to \rqn{1}}: Overall, practitioners perceived TagDebt as a useful tool to support TDM: most respondents (13/16) highlighted that it increases the visibility of TD, while a considerable number (9/16) reported that it helps increase awareness and supports prioritizing TD items. However, themes such as classification precision (mentioned by 10/16 respondents) and the lack of more descriptive labels and explanations (mentioned by 6/16 respondents) are pointed out as limitations that could hinder trust in the bot.

\end{findingbox}

\subsection{Ease of Use of TagDebt}

To answer \rqn{2}, we investigated how practitioners perceive the ease of use of the TagDebt bot. Overall, participants discussed several aspects related to learning, configuring, and operating the bot in their workflows. The topics related to ease of use were classified into three categories: (i) \textit{Installing and Configuring}; (ii) \textit{Learning and Understanding}; and (iii) \textit{Operating}. Table~\ref{tab:themes-ease-of-use} shows the themes within each category.

\begin{table}[h]
\caption{Themes related to ease of use, the respondents supporting each theme, and the total of indicators per theme}
\label{tab:themes-ease-of-use}
\begin{tabular}{
p{4cm}
p{0.1cm}
p{0.1cm}
p{0.1cm}
p{0.1cm}
p{0.1cm}
p{0.1cm}
p{0.1cm}
p{0.1cm}
p{0.1cm}
p{0.2cm}
p{0.2cm}
p{0.2cm}
p{0.2cm}
p{0.2cm}
p{0.2cm}
p{0.2cm}
p{0.1cm}}
\hline
\multicolumn{1}{c}{\multirow{2}{*}{\textbf{Themes}}} & \multicolumn{16}{c}{\textbf{Respondents}} \\
\multicolumn{1}{c}{} & \textbf{r1} & \textbf{r2} & \textbf{r3} & \textbf{r4} & \textbf{r5} & \textbf{r6} & \textbf{r7} & \textbf{r8} & \textbf{r9} & \textbf{r10} & \textbf{r11} & \textbf{r12} & \textbf{r13} & \textbf{r14} & \textbf{r15} & \textbf{r16} & \textbf{Ind.} \\ \hline
\vspace{0.1cm}

\textbf{Installing and Configuring} &  &  &  &  &  &  &  &  &  &  &  &  &  &  &  & & \\
Simple to install, configure, and operate & \checkmark & \checkmark &  & \checkmark & \checkmark & \checkmark & \checkmark &  & \checkmark & \checkmark & \checkmark & \checkmark & \checkmark & \checkmark &  & \checkmark & 27 \\

Configuration file is clear & \checkmark &  &  & \checkmark &  &  &  & \checkmark & \checkmark &  & \checkmark &  &  & \checkmark & \checkmark & \checkmark & 10 \\

\vspace{0.1cm}
\textbf{Learning and Understanding} &  &  &  &  &  &  &  &  &  &  &  &  &  &  &  &  \\
Documentation is simple and complete & \checkmark & \checkmark &  & \checkmark & \checkmark & \checkmark & \checkmark & \checkmark & \checkmark & \checkmark & \checkmark & \checkmark & \checkmark & \checkmark & \checkmark & \checkmark & 28 \\

Documentation has some limitations &  &  & \checkmark & \checkmark &  &  &  &  & \checkmark & \checkmark &  &  &  & & &  & 7 \\

Remembering how to use the bot & \checkmark & \checkmark & \checkmark & \checkmark & \checkmark &  & \checkmark & \checkmark & \checkmark & \checkmark &  &  & \checkmark & \checkmark & \checkmark & \checkmark & 14 \\

\vspace{0.1cm}
\textbf{Operating} &  &  &  &  &  &  &  &  &  &  &  &  &  &  &  &  \\
Commands as comments are not ideal &  &  &  &  & \checkmark &  &  & \checkmark &  & \checkmark &  &  &  &  & \checkmark &  & 4 \\ \hline

\end{tabular}%
\end{table}

TagDebt was considered \textbf{Simple to install, configure, and operate} (supported by 13/16 respondents; 27 indicators): \indicator{5}{It was quite simple. I had no kind of issue during the installation process.} Installing it was considered straightforward, which was expected due to the integration of the bot with GitHub: \indicator{9}{I managed to install it in the repository, create the file, and classify the issues in less than 10 minutes}. Similarly, practitioners considered that the \textbf{Configuration file is clear} (supported by 8/16 respondents; 10 indicators), which indicates that JSON configuration files are familiar to most developers, and may help to reduce the overhead for configuring the bot: \indicator{1}{I think the JSON file is fine}. However, a participant mentioned that the number of configuration options might be overwhelming: \indicator{9}{It’s confusing also in the configuration part of the JSON file… I think since there are so many configurations, it ends up being a bit confusing.} In this scenario, practitioners recognize that the default \filename{config.json} provided is handy and helps to reduce the time and effort to configure the bot: \indicator{9}{You kind of just go with the default configuration file, you know? Just to start using it quickly and not waste too much time.} 

In a second category, \textbf{\underline{Learning and Understanding}}, we identified that the \textbf{Documentation is simple and complete} (supported by 15/16 respondents; 28 indicators), indicating that practitioners found TagDebt's documentation clear, straightforward, and well-structured. The quick-start guide and configuration guides were noted for making the setup easier: \indicator{9}{Yes, I found it [the quick-start guide] really easy to find my way around the sections. I think it is well structured.} There was an example of a configuration file, which also contributed positively to the ease of use of the documentation: \indicator{9}{I really liked that there’s a separate repository with the configuration and the classification the bot gave to each issue… it really helped to understand.} However, the clarity of the documentation appears to be tied to the technical experience of the user: \indicator{1}{You have to have a programming background... developers will understand very easily… no one else. Neither would a technical PM probably understand the whole thing.} This suggests that while the documentation is effective for technical staff, managers might consider it less clear. Some suggestions for improvement emerged, such as including more visual references in the description of the marketplace: \indicator{2}{Maybe a screenshot directly in the documentation [on GitHub Marketplace] could help, because we have to open another link [to access the quick-start guide]}.

Although the documentation was considered clear, we identified some flaws reported by practitioners, as per the theme \textbf{Documentation has some limitations} (supported by 4/16 respondents; 7 indicators). A recurring issue was a lack of clarity about how the bot works and what users should expect during installation and usage: \indicator{3}{It wasn't clear at all which one was supposed to do, because one hand mentions you can install from marketplace and on the other it mentions that if something is not working, you have to install it properly... It is a bit involved, to be honest.} There was also uncertainty about configuration details: \indicator{3}{Even though I installed it from the marketplace, I was really confused about configuration. I was expected to find something to tell what labels to use. But the configuration did not tell me what the defaults are... so I assumed it had some defaults, but it didn’t contribute to the positive experience.} Others reported taking longer than expected to understand how the bot classifies the issues: \indicator{10}{I liked it a lot, but it took me a bit to understand how it would work… I thought it would do some automatic mapping or scanning in the code… then I understood it was based on GitHub issues.}

Another theme that emerged from the data analysis was \textbf{Remembering how to use the bot} (supported by 13/16 respondents; 14 indicators). For this theme, we identified mixed feelings among the participants. Most of them found the bot commands relatively simple to remember, and they would not face problems in using the bot: \indicator{1}{all the parameters in common are quite easy to remember} and \indicator{8}{there are relatively few commands and there’s a help command. I think it’s very straightforward.} On the other hand, some participants highlighted that the bot has many configuration options, making it difficult to remember how to use it after a period of inactivity: \indicator{7}{If I don’t use it like every day, I will forget the command, I think} and \indicator{9}{I think I might have difficulty, not because it's hard, but because we have a lot on our minds.} Nonetheless, the participants also mentioned that the help command is a very useful mechanism to help them remember how to use the bot: \indicator{2}{Once a week, I would have to use the help because I forgot how to auto-categorize or auto-label... but after using the help, it’s fine.} and \indicator{10}{I remember that there’s a help command that shows everything, so I’d manage.}

In the category \textbf{\underline{Operating}}, we found the theme \textbf{Commands as comments are not ideal} (supported by 4/16 respondents; 4 indicators). One participant mentioned that keeping commands as comments increases the size of issues and may hinder practitioners' understanding: \indicator{5}{If I write a comment like label, I mean, it's not interesting for someone else to know it… Why should I have a persistent comment that is just a comment?} Additionally, there was a suggestion that the bot’s interaction model (through comments) should be more explicitly documented: \indicator{8}{Maybe the documentation should state that the interaction happens through comments.} Another participant mentioned that it was not clear that the bot would not put a comment with the classification: \indicator{10}{One thing I didn’t understand well was when I put the command to add a label, it just said it classified or not, but I didn’t see the classification. I was like, Where is the classification?}

\begin{findingbox}
    \textbf{Answer to \rqn{2}}: Overall, practitioners found TagDebt easy to install, configure, and operate. Most respondents (13/16) indicated that the bot is simple to install, configure, and use, and half of the respondents (8/16) mentioned that the configuration file is clear and familiar. In contrast, some respondents indicated concerns about understanding the configuration details (4/16) and using comments as input for the bot (4/16), highlighting that there is still room for polishing and improving the usability of the TagDebt bot.
\end{findingbox}

\subsection{Contextual Factors for Using TagDebt}
\label{sec:results-contextual-factors}

To answer \rqn{3}, we investigated the contextual factors that influence practitioners’ intention to use the TagDebt bot. Contextual factors refer to aspects that go beyond the tool’s internal features (e.g., team characteristics or project environment) but still shape its adoption and perceived value. The topics related to contextual factors were grouped into five themes, which we summarize in Table~\ref{tab:themes-contextual-factors}.

\begin{table}[h]
\caption{Contextual factors that might influence the adoption of TagDebt, the respondents supporting each factor, and the total of indicators per factor}
\label{tab:themes-contextual-factors}
\begin{tabular}{
p{4cm}
p{0.1cm}
p{0.1cm}
p{0.1cm}
p{0.1cm}
p{0.1cm}
p{0.1cm}
p{0.1cm}
p{0.1cm}
p{0.1cm}
p{0.2cm}
p{0.2cm}
p{0.2cm}
p{0.2cm}
p{0.2cm}
p{0.2cm}
p{0.2cm}
p{0.1cm}}
\hline
\multicolumn{1}{c}{\multirow{2}{*}{\textbf{Themes}}} & \multicolumn{16}{c}{\textbf{Respondents}} \\
\multicolumn{1}{c}{} & \textbf{r1} & \textbf{r2} & \textbf{r3} & \textbf{r4} & \textbf{r5} & \textbf{r6} & \textbf{r7} & \textbf{r8} & \textbf{r9} & \textbf{r10} & \textbf{r11} & \textbf{r12} & \textbf{r13} & \textbf{r14} & \textbf{r15} & \textbf{r16} & \textbf{Ind.} \\ \hline

Team Size & \checkmark &  & \checkmark &  &  &  & \checkmark &  &  &  &  &  & \checkmark & \checkmark & \checkmark &   & 12 \\

Code Base Size &  &  & \checkmark &  & \checkmark &  & \checkmark &  &  &  &  & \checkmark &  & \checkmark & \checkmark &  & 8 \\

Practitioners’ Roles & \checkmark &  & \checkmark & \checkmark &  &  &  &  &  &  & \checkmark &  &  & \checkmark & & & 7 \\

Privacy and Security Policies & \checkmark &  &  &  &  &  &  & \checkmark &  &  &  &  &  &  &  &  & 3 \\

Type of software &  &  &  &  &  &  &  & \checkmark &  &  &  &  &  &  &  & \checkmark & 3 \\ \hline

\end{tabular}%
\end{table}

First, participants suggested that \textbf{Team Size} (supported by 6/16 respondents; 12 indicators) impacts the decision to adopt TagDebt. Several interviewees indicated that the benefits of using the bot become more evident when the number of team members increases: \indicator{3}{[I think the bot was] designed for people who work in large teams}. This happens especially because there might be bottlenecks in labeling when multiple developers are opening issues in the same repository. However, the bot would also not be ideal if there are many contributors/team members. According to the practitioners, the bot would be mostly useful for teams between 10 and 25 developers/contributors: \indicator{3}{Oh well, I'm not sure these are the correct terms to talk about teams. I mean, a single person can manage only this many people, so are 10 people.} and \indicator{1}{But also with some limitations at some points, only the bot will not be enough. If you have a team of 25 people, commit.} On the other hand, in small teams, the bot seems to be irrelevant. One participant highlighted that they would not even consider using the bot if only one or two people are working in a given project: \indicator{3}{I found it quite useless for my project [...] no one else collaborates, so there is no need to communicate.} While this suggests a limitation in perceived applicability for smaller teams, it also reinforces the bot’s potential value in larger, more collaborative settings.

Practitioners also mention that in smaller projects, the manual effort required to label issues is minimal, and therefore the \textbf{Codebase Size} (supported by 6/16 respondents; 8 indicators) is a factor to consider when adopting TagDebt: \indicator{3}{the amount of tasks they'd have to classify has to be quite high because it doesn't take much effort to classify a task and this doesn't have to be done so often.} However, in contexts with a high volume of issues, participants recognized the potential for errors and oversights, such as developers forgetting to assign labels, which is the context where TagDebt seems suitable: \indicator{5}{with the increase of numbers, the importance of the bot increases. With huge numbers, it makes sense to automate the classification}.

\textbf{Practitioners' Roles} (supported by 5/16 respondents; 7 indicators) is a theme indicating that different types of stakeholders might not be interested in adopting the bot. For instance, one participant described how the adoption of the tool is typically a request that comes from developers, who are more interested in using the bot and are often responsible for introducing the new tools: \indicator{1}{The bot is very useful. This is how it works, and they will tell this to their PM team, leads product manager, whoever, OK. And the developers need to clarify the intention of the bot.} In addition, different roles within the team perceive the bot’s functionality differently. Some developers do not consider labeling essential: \indicator{3}{for some people, it doesn't matter what it [the bot] says}. However, participants mentioned that managers could benefit from using the bot if they are responsible for managing issues and backlogs: \indicator{10}{Generally, it is more relevant for the agility person, who handles things like the Scrum board, the Kanban, etc., within the team. It's usually this person who looks at what is being done and what is being addressed. What often happens in a more technical context is that, in a company that has a chapter or something more organized among developers of the same area, this comes into play.}.

In the \textbf{Privacy and Security Policies} (supported by 2/16 respondents; 3 indicators) theme, we summarize that participants have concerns about using the bot in organizations where code privacy and data control are critical: \indicator{1}{most of the time you don't want to give access to any third-party solution to your private repository}. This concern is mainly related to critical domain applications, such as finance and health insurance: \indicator{1}{for banking, insurance, and all the finance, they will not push any code to GitHub.com}. In this scenario, practitioners mentioned that organizations would prefer to have their own deployed instance of the bot, which is a possible solution: \indicator{1}{I imagine the organizations that I met don’t like GitHub but [prefer] Enterprise, GitLab, and all those kinds of solutions}. 

Finally, the \textbf{Type of Software} (supported by 2/16 respondents; 3 indicators) (i.e., proprietary or OSS) also may impact the decision to adopt TagDebt. The practitioners mentioned that the bot would benefit OSS projects more: \indicator{7}{Uh, I think the issue is more like an OSS project use case. Because there are going to be other people who will do it. Request or demand the issue.} Hence, while the bot can be helpful in both settings (as already highlighted), internal policies and workflows could impact the adoption of the bot in private organizations. This shows that developers working in OSS projects would be the primary audience for TagDebt.

Practitioners indicated that the intention to use TagDebt depends on several contextual factors, such as the codebase size and team size: while small projects or teams find little value in automation, larger and more collaborative settings see greater benefits in reducing errors and managing workload. The roles of practitioners also influence the adoption, as developers often introduce the tool, but managers or roles responsible for backlogs and boards may gain more from its outputs. Concerns about privacy and security policies can limit adoption in domains like finance and health, where self-hosted instances are preferred over third-party services. Finally, the type of software matters, with open source projects perceived as a more natural fit due to their collaborative nature, while private organizations may face restrictions from internal policies.

\begin{findingbox}
    \textbf{Answer to \rqn{3}}: Overall, practitioners indicated that the intention to use TagDebt depends on several contextual factors, such as \textbf{Team size} (6/16 respondents) and \textbf{codebase size} (6/16 respondents). Specifically, small projects or teams may see little value in automation, while larger and more collaborative settings benefit more from using the bot. Besides, \textbf{Privacy and security policies} were explicitly mentioned by 2/16 practitioners (3 indicators), indicating that in sensitive domains (e.g., finance, insurance), self-hosted instances of TagDebt might be preferred.
\end{findingbox}

\subsection{Improvements for TagDebt}
\label{sec:results-improvements}

To answer \rqn{4}, we identified the improvements to the bot suggested by the respondents. Overall, the suggestions cover both technical and usability aspects, such as a more intuitive configuration process, integration with other existing tools, and better communication features. Table~\ref{tab:themes-improvements} summarizes the list of improvements we identified.

\begin{table}[h]

\caption{List of improvements for TagDebt, the respondents supporting each theme, and the total of indicators per theme}
\label{tab:themes-improvements}
\begin{tabular}{
p{4cm}
p{0.1cm}
p{0.1cm}
p{0.1cm}
p{0.1cm}
p{0.1cm}
p{0.1cm}
p{0.1cm}
p{0.1cm}
p{0.1cm}
p{0.2cm}
p{0.2cm}
p{0.2cm}
p{0.2cm}
p{0.2cm}
p{0.2cm}
p{0.2cm}
p{0.1cm}}
\hline
\multicolumn{1}{c}{\multirow{2}{*}{\textbf{Themes}}} & \multicolumn{16}{c}{\textbf{Respondents}} \\
\multicolumn{1}{c}{} & \textbf{r1} & \textbf{r2} & \textbf{r3} & \textbf{r4} & \textbf{r5} & \textbf{r6} & \textbf{r7} & \textbf{r8} & \textbf{r9} & \textbf{r10} & \textbf{r11} & \textbf{r12} & \textbf{r13} & \textbf{r14} & \textbf{r15} & \textbf{r16} & \textbf{Ind.} \\ \hline

Integration and Analysis & \checkmark & \checkmark & \checkmark & \checkmark &  &  &  & \checkmark &  & \checkmark &  & \checkmark & \checkmark &  & \checkmark & \checkmark & 28 \\

Feedback for users & \checkmark & \checkmark &  &  &  &  &  & \checkmark &  & \checkmark &  & \checkmark &  &  & \checkmark & \checkmark & 9 \\

Label management & \checkmark &  & \checkmark &  & \checkmark &  &  &  & \checkmark &  & \checkmark &  &  &  & &  & 9 \\

Configuration and Setup &  &  &  &  & \checkmark & \checkmark &  &  & \checkmark &  & \checkmark &  & \checkmark &  & &  & 7 \\ \hline

\end{tabular}%
\end{table}

Participants suggested \textbf{Improvements in integration and analysis} (supported by 10/16 respondents; 28 indicators) and suggested that the bot could use \textbf{multiple data sources}, such as the full issue discussion, source code, design documents, meetings, and other technical artifacts, to enhance the explainability of issue classification. There was also a suggestion to \textbf{integrate the bot with GitHub Copilot}, which is also aligned with leveraging more data sources and would enable the bot to consider source code in the issue classification. Finally, the bot could further help issue monitoring if the \textbf{issue complexity} were also considered. Nonetheless, it was not possible to identify how to measure the complexity.

Regarding the interaction with practitioners, \textbf{Improvements in communication and feedback} (supported by 7/16 respondents; 9 indicators) were also mentioned. Users expect the bot to provide \textbf{more actionable and contextual feedback}, especially during the development workflow. For instance, \textbf{providing comments related to the issue directly in the code} further helps practitioners to contextualize the TD items. Participants also mentioned the bot could \textbf{suggest how to fix a problem}, providing more information for practitioners' decision-making. Finally, participants also mentioned that the bot could also be useful if used in \textbf{pull requests}, where discussions about TD also happen. These features would make the bot more useful in critical stages of the software life cycle.

Improvements in \textbf{Label management} (supported by 5/16 respondents; 9 indicators) were also suggested by practitioners. They suggested that the bot should \textbf{reuse existing repository labels}, ensuring standardization and better alignment with current workflows. The need to \textbf{reconcile multiple labels} applied manually or automatically was also discussed, with options for customization. Another important request was the ability to \textbf{classify the issue by the module or part of the system} it relates to, giving more contextual information to the assigned label. Finally, users reported that the bot could also help identify \textbf{similar or duplicated issues}, especially in projects with a long history of issues. This improvement facilitates the visualization of issues, helping in their management.

Practitioners suggested that we carry out \textbf{Improvements related to configuration and setup} (supported by 5/16 respondents; 7 indicators), such as making the bot’s configuration process more practical and centralized, for instance, by having a \textbf{global configuration file} for multiple repositories and/or providing \textbf{support for maintaining multiple configuration files}, which will be used in multiple repositories. Furthermore, participants suggested that we provide \textbf{an external page with a graphical interface} for generating and customizing settings. This would reduce the effort required to understand the configuration options. Finally, it was suggested that \textbf{commands issued as comments} could be automatically deleted after execution to keep the discussion cleaner.

\begin{findingbox}
    \textbf{Answer to \rqn{4}}: Overall, \textbf{Integration and analysis} were the most frequently mentioned improvements (10/16 respondents). Specifically, practitioners called for richer integration and analysis capabilities, such as using multiple data sources (issue discussions, code, documents), considering issue complexity, and integrating with tools like GitHub Copilot. Regarding \textbf{Configuration and setup}, more centralized configuration (e.g., reusable or global configuration files) was required. Finally, 7/16 respondents suggested that the bot provide more contextual and actionable support, including comments linked to code, suggestions for fixes, and support in pull requests.
\end{findingbox}

\section{Discussion}
\label{sec:discussion}

This study was motivated by two research problems: \textit{``(i) Is it possible to develop a specialized TDM tool that is easy to integrate within existing software development workflows?''}; and \textit{``(ii) Considering that we can develop such a tool, would practitioners intend to adopt it?''}. After developing and evaluating TagDebt, we learned that bots are a good alternative for seamlessly including TDM in existing workflows and increasing the perceived value of TDM. Our results also reveal that, beyond technical integration, factors such as trust, explainability, and ease of configuration play an equally important role in determining the actual usefulness of a TDM tool. Hence, while integrating it into workflows is possible, the tool design must account for socio-technical dimensions, ensuring that automation complements developers' reasoning processes.

Regarding the adoption of TDM tools, the context plays a crucial role, particularly in terms of team size, organizational alignment, and regulatory constraints. Larger teams seem to benefit significantly more from TDM tools like TagDebt, as they alleviate coordination and labeling bottlenecks, whereas smaller teams or individual developers may see limited value. Moreover, adoption is facilitated when tools demonstrate transparency in their operations and provide clear, simple documentation. In the following sections, we elaborate on the evaluation results to illustrate these findings in detail. 

\subsection{Interpretation of the Results of \rqn{1}}
\label{sec:intepret-rq1}

According to practitioners, the \textbf{usefulness} of TagDebt has, to some extent, a socio-technical aspect. As for most TDM tools, TagDebt can achieve broader goals, such as improving team coordination and improving communication among the team. In addition, TagDebt complements developers’ own reasoning processes, instead of replacing and fully automating TD identification. TagDebt's ability to increase visibility and awareness of TD supports better decision-making and may lead the teams to better define priorities and reduce friction in communication.

Another aspect related to its usefulness is trust in automation. Although participants acknowledged productivity improvements, transparency and explainability are mentioned as key to validating bot decisions. This information is aligned with existing literature \citep{Biazotto2025, Biazotto2025b, Avgeriou2025}, which shows the need to keep humans in the loop when managing TD. TagDebt's evaluation shows that the value of TDM tooling is not only related to the tool's functionality, but also to the interaction between developers and the tool. To maximize adoption, bots like TagDebt must offer more than correct classifications: they must foster trust, seamlessly integrate into existing practices and workflows, and contribute to shared understanding within teams.

Overall, our findings shed light on how developers perceive workflow-integrated automated assistants in their daily work. Practitioners usually perceive TagDebt as a way to reduce manual effort (e.g., avoiding repetitive labeling) and increase visibility and awareness of TD, but they consistently framed the bot as a supporting tool rather than a fully autonomous tool. Concerns about classification precision and the demand for more descriptive labels and explanations indicate that developers rely on the bot as a starting point for analysis or “second opinion,” but they still expect to keep control over final decisions. This suggests that, for TD-related automation, transparent feedback (e.g., why a label was assigned, which parts of the issue triggered the classification) is key for tool adoption.

\subsection{Interpretation of the Results of \rqn{2}}
\label{sec:intepret-rq2}

The findings for RQ2 indicate that TagDebt is \textbf{easy to use}, especially due to its smooth integration with GitHub, and having a centralized configuration file for each repository. These attributes suggest that other TDM tools could focus on effectively aligning with common development workflows and developers' preferences, such as reusing familiar configuration practices (e.g., JSON/YAML). Therefore, TagDebt might have a low barrier to adoption.

While the overall user experience seems to be good, a few issues were identified and therefore represent opportunities to refine the bot and its documentation. First, while the number of options in the configuration file increases the developer's control over the bot (which is aligned with the results we found in \citep{Biazotto2025b}), it may become slightly convoluted for first-time users. This issue may also be the reason some practitioners acknowledge that they would not remember the configuration options if they do not use the bot on a daily basis. Hence, when developing tools for TDM, refining the configuration files and specifying a minimum set of options are good practices to reduce the complexity in configuring the tools. Similarly, while the documentation was considered clear, some participants had issues understanding the purpose and classification mechanisms of the bot (e.g., it was not clear that the source code would not be analyzed). Then, new TDM tools should have simple and straightforward documentation to avoid such issues and improve their adoption. 

The reported ease of installation, configuration, and use (e.g., most respondents considered the documentation simple and the configuration file clear) suggests that practitioners are open to adopting lightweight, workflow-integrated automation for TD, especially when it leverages familiar tooling (GitHub) and requires minimal onboarding. At the same time, our data indicate that this readiness is, in part, due to TagDebt's simplicity (i.e., it supports only issue labeling and notifications), with limited configuration overhead and a clear interaction model. Although more sophisticated AI-based TDM tools may offer richer analyses (e.g., cross-artifact reasoning over issues, code, and architecture), they will likely need to preserve this low-overhead integration while also providing more substantial support for understanding and controlling automated decisions. Hence, our results suggest that practitioners are ready for more AI-driven TDM support, as long as it remains transparent, lightweight, and well-aligned with existing workflows.

\subsection{Interpretation of the Results of \rqn{3}}
\label{sec:intepret-rq3}

In RQ3, we identified that the \textbf{intention to adopt and use} TagDebt is influenced by several \textbf{contextual factors}. One of the most cited factors is \textit{team size} (which reinforces the socio-technical aspect, as mentioned in Section~\ref{sec:intepret-rq1}). Practitioners consistently stated that larger teams would benefit more from adopting TagDebt. When these teams start to work on repositories and open issues, it is common for labeling bottlenecks to occur. In addition, coordinating and standardizing labeling on those larger teams becomes challenging, and therefore, it would make sense to adopt the bot. In contrast, for smaller teams or individual developers, the bot seems to be unnecessary. This is because the number of issues and labels is too low, and developers would spend more time reviewing the bot's suggestions than adding the labels themselves. Therefore, when developing tools for TDM, scoping the environment for tool adoption is crucial to extract more value from them.

Similarly to the previous factor, the developer's role in this context also impacts the adoption of TagDebt. In general, developers are the main audience for the bot, although managers can benefit from using the bot if they are responsible for managing and prioritizing the issues. This factor reinforces the importance of \textit{organizational alignment} in tool adoption: while developers may benefit immediately, the tool's success also depends on management expectations and the perceived value in the long term. 

\textit{Privacy and infrastructure constraints} emerged as another key factor. Relying on GitHub infrastructure may raise concerns for companies in sectors where regulatory or security constraints are stricter (e.g., finance and insurance). Hence, our decision to make the bot open source, enabling local deployment and support, may help work around privacy constraints.

These findings indicate that the decision to adopt TagDebt, and possibly other TDM tools, is highly contextual. Therefore, when proposing tools for TDM, it is crucial to incorporate adaptability and configurability as core design principles to better support adoption across multiple environments.

The contextual factors we identified (e.g., team size, codebase size, roles, privacy and security policies, and software type) highlight that adoption decisions for workflow-based TDM tools are inherently socio-technical. Many practitioners perceived TagDebt as particularly valuable in medium-to-large, collaborative teams and projects with a substantial number of issues, where manual labeling becomes error-prone and time-consuming. Conversely, in very small teams or small codebases, automation was often seen as unnecessary overhead. Additionally, strict privacy and security policies in certain domains (e.g., finance, health) constrain the use of third-party services, motivating self-hosted or on-premise deployments. Together, these findings emphasize that tool adoption cannot be evaluated in isolation from team structure, governance, and organizational constraints.

\subsection{Interpretation of the Results of \rqn{4}}
\label{sec:intepret-rq4}

Some improvements to TagDebt involve streamlining the configuration and setup process, which reinforces the idea that TDM tools must introduce minimal friction to be adopted. Alternatives to reduce friction include centralized configuration management and making the configuration process easier, for instance, by using a graphical interface. Hence, when developing tools for TDM, vendors should take into account that the tools' operational complexity is a potential barrier for adoption.

Beyond ease of use, practitioners clearly desire more \textit{context-aware and integrated analytical capabilities}. The ability to draw insights from multiple sources (ranging from issue discussions to source code) and from design documentation was seen as essential for improving the precision and relevance of TD classification. This reflects that effective TDM requires understanding technical artifacts holistically rather than relying on isolated data points.

In addition, users indicated that the bot could engage more proactively within development workflows, such as suggesting refactoring actions on pull requests. This preference implies that TDM tools should provide information that can be easily checked by practitioners, and lead to practical actions.

Labeling and organizational features also garnered attention, with calls for better reuse of existing labels and semantic enrichment through module-based classification. These improvements would foster alignment with existing team conventions and improve the interpretability of debt items, ultimately supporting prioritization and strategic planning.

The themes that emerged from our study suggest several design lessons for other workflow-integrated tools. First, embedding functionality directly into existing artifacts (e.g., GitHub issues) and reusing project concepts (e.g., repository labels) can reduce the cognitive and process overhead for tool adoption. Second, providing sensible defaults (e.g., a ready-to-use configuration file) while allowing gradual customization helps teams get started quickly and invest further effort only if the tool proves valuable. Third, interaction mechanisms should be designed to balance simplicity with cleanliness of the workflow. For instance, participants appreciated comment-based commands but also requested automatic removal of “command” comments to avoid cluttering issue discussions. Finally, our results underline the importance of explainability and richer feedback (e.g., justifications for labels, pointers to relevant parts of the issue, integration with code review or pull requests). These aspects can inform the design of future workflow-based TDM tools and, more broadly, other bots that aim to support developers without disrupting established practices.

Overall, the suggested improvements reflect a desire for TagDebt to evolve from a straightforward labeling tool into a more \textit{integrated, intelligent assistant} that balances usability with deeper contextual understanding. This evolution could make the bot more relevant across diverse team sizes, project complexities, and development contexts, ultimately fostering more effective TDM practices.

\subsection{Implications for Practitioners}
\label{sec:impl-practitioners}

The findings from our study on the TagDebt bot offer several valuable insights for practitioners and tool vendors. These findings are associated mainly with the adoption or improvement of automated tools for TDM. Below, we summarize each of those implications:

\begin{enumerate}
    \item \textit{Enhance Visibility and Prioritization of TD Items:} Practitioners can significantly benefit from automating the classification of SATD items by using TagDebt. The bot helps teams maintain the visibility of debt. This is particularly critical in larger teams or projects where manual labeling becomes impractical. Incorporating such automation can reduce the cognitive load on developers and minimize errors caused by manual labeling or oversight.

    \item \textit{Team and Environment Characteristics when adopting TDM tools:} We help practitioners understand how the environment and the team impact the perceived usefulness of automated TDM tools, which is highly context-dependent. For instance, larger teams benefit from tools that help coordinate actions. In contrast, smaller teams may find limited value, suggesting that adoption decisions should be carefully considered. Furthermore, practitioners should be aware of privacy and security policies, especially in sensitive or regulated environments, ensuring that tool integrations comply with organizational constraints.
    
    \item \textit{Balancing Ease of Use with Configuration Flexibility:} While TagDebt is considered easy to install and operate, practitioners recognize that it has an initial learning curve related to configuration and command usage. Hence, vendors should be aware that a higher number of configuration options may lead to complexity in setting up the tools. Therefore, effective onboarding, supported by clear documentation and ready-to-go setups, is essential to facilitate adoption and continued use of TDM tools.
    
    \item \textit{Integrating Tools Seamlessly into Developer Workflows:} To maximize tools' value, vendors should provide functionality that works directly within the development life cycle, such as comments in pull requests or code suggestions. This tight integration improves the relevance of TD information and may increase the awareness of TD.
    
    \item \textit{Continuous Improvement through User Feedback:} Our study highlights the importance of engaging with users to identify improvements to existing tools. Our study also advocates for more explainability and transparency in tools to improve trust in them. Vendors should detail how data is analyzed to help practitioners understand how to use the information.

    \item \textit{Low-Overhead and Explainable Tools:} We deem that such results suggest similar principles apply to workflow-integrated tools in general. For developers to perceive a tool as genuinely useful, it should require minimal configuration and setup effort, integrating into existing workflows with little friction rather than demanding extensive tuning or process changes. At the same time, tools that classify, recommend, or prioritize items (e.g., refactoring actions) need to provide explainable outputs, making clear why a decision was made, which signals were used, and with what level of confidence. This would help practitioners quickly validate, contest, or refine those decisions.

    \item \textit{Trade-offs in tool usage:} These findings can also be transferred to tool usage more broadly. Our results indicate that adopting tools usually leads to a trade-off between configuration effort and learning curve on one side, and expected productivity gains on the other. Teams are more willing to invest in setting up and learning a tool when system size or process complexity is high enough that manual work becomes a clear bottleneck. In contrast, in small teams or simpler projects, the perceived overhead of configuring, integrating, and mastering a new tool can hinder its benefits, leading practitioners to prefer manual solutions. Therefore, when designing and evaluating development tools in general, it is essential to consider where this balance lies in different contexts and to minimize upfront overhead so that productivity benefits become visible early in the adoption process.
\end{enumerate}

In summary, practitioners who aim to implement or optimize TDM tools such as TagDebt should prioritize automation that enhances visibility and prioritization, carefully consider their team and project context, emphasize usability and seamless workflow integration, and actively engage in continuous improvement to maximize the benefits of such technologies.

\subsection{Implications for Researchers}
\label{sec:impl-researchers}

The findings from our investigation into the use and perception of TagDebt provide several directions and considerations for researchers working on TDM tools and SE automation.

\begin{enumerate}
    \item \textit{Exploring Contextual Factors in Tool Adoption:} Our results highlight the critical role of contextual factors (such as team size, developer roles, project type, and organizational privacy constraints) in shaping the adoption and perceived usefulness of TDM tools. Researchers should investigate these contextual dimensions more deeply, potentially developing frameworks or models that predict when and how such tools are most beneficial. This understanding can guide the design of adaptable tools that fit diverse development environments.
    
    \item \textit{Advancing Usability Research for Developer Tools:} While ease of use was generally acknowledged, challenges remain around configuration complexity, remembering commands, and interaction modalities (e.g., commands via comments). Future research should focus on improving the usability of automated tools through better interfaces, contextual help, and more intuitive interaction models. Studies involving longitudinal user experience and cognitive load can provide valuable insights into sustained tool adoption.
    
    \item \textit{Enhancing Automated Analysis and Feedback Mechanisms:} Participants expressed a desire for richer analysis that incorporates multiple data sources beyond issue titles, such as code, documentation, and discussion threads, to improve classification accuracy and contextual relevance. Researchers can explore novel techniques in natural language processing, machine learning, and software repository mining to develop more sophisticated models to identify and prioritize TD. In addition, embedding actionable feedback directly into developers’ workflows remains an open challenge that deserves further exploration.
    
    \item \textit{Facilitating Continuous Improvement and User-Centered Design:} The diverse suggestions for feature enhancements emphasize the importance of iterative, user-centered design processes in the development of TDM tools. Researchers should work closely with practitioners to identify evolving needs and assess the impact of new features. Mixed-method studies that combine quantitative usage metrics with qualitative feedback can help bridge the gap between tool capabilities and practical utility.
\end{enumerate}

Overall, this study underscores several promising avenues for advancing research on automated TDM tools. By addressing contextual influences, improving usability, expanding analytical depth, and fostering collaboration with practitioners, researchers can contribute to more effective and widely adopted solutions in SE.

\section{Threats to Validity}
\label{sec:tov}

As in any empirical study, there are some threats to the validity of our results and contributions. We deemed that each phase of our research method (DSR) may pose specific threats. To ensure that we cover all potential threats and provide a better understanding of how those threats spread throughout the research method, we organize this section considering the DSR phases (i.e., Understand the social context, Understand the knowledge context, Design, and Investigation) and also discuss the associated mitigation actions.

In the \textbf{Understand the social context} and \textbf{Understand the knowledge context} phases, our objective was to understand the problems and demands with respect to TDM and the potential solutions to such problems. Both phases suffer from the same threat: since we did not apply a systematic approach for reviewing the literature (e.g., SMS or systematic literature review), it is possible that some problems and/or solutions related to TDM were not identified. However, we mitigated those threats by focusing on recent literature, which we considered to contain the state of the art on TDM. For instance, the Manifesto on Reframing Technical Debt, a core reference in our study, is the result of a seminar and focus groups in which the most prominent researchers in TD were involved, and therefore, it summarizes the current understanding of TDM needs. In addition, our two previous studies \citep{Biazotto2025,Biazotto2025b} highlight the needs of practitioners from two different sources (i.e., StackExchange posts and a survey with practitioners), providing strong evidence on the need for more specialized TDM tools. Our SMS (\cite{Biazotto2023}) summarizes information on more than 120 TDM tools and provides a broad understanding of the tooling for TDM. Focusing on recent evidence, we are confident that we have a good overview of current TDM problems and solutions.

Regarding the \textbf{Design} phase, we addressed a solution to TDM problems. As mentioned in Section~\ref{sec:background}, a core challenge faced by developers is the friction that TDM tools usually impose on existing development workflows. Our suggestion of proposing a bot, rather than other types of tools, is also a potential threat. We mitigate this threat by basing our decision on previous work that showed that bots are widely accepted by developers and, therefore, bots could be a good alternative to support TDM~\citep{Phaithoon2021, Biazotto2023}. Another threat within the Design phase is the adoption of an NLP solution. Although current research on NLP solutions for TD detection showed their potential, practitioners still do not fully trust such solutions, which could affect their perceptions about the bot. However, we deemed that any type of tool would face similar challenges, as presented in \citep{Biazotto2025}. To mitigate this threat, we focused on a very specific task (i.e., assigning labels), which can be easily reviewed by practitioners. In addition, we simplified model replacement, which might help practitioners keep up-to-date with advances in AI for TDM. 

Another threat to validity is that the requirements implemented in TagDebt might not be ideal or complete for all contexts in which practitioners manage TD. We reused a previously published set of practitioner-derived requirements from our prior work~\citep{Biazotto2025b}. While this mitigates the risk of ad-hoc requirement definition by grounding the tool design on empirical evidence, it also implies that our selection of a subset of requirements may not capture all expectations that different teams may have for TDM tooling. We mitigate this threat in two ways: first, we selected requirements that together compose a coherent end-to-end use case (i.e., issue labeling complemented by configurable notifications), rather than implementing isolated features; second, we evaluated the resulting artifact with practitioners to elicit improvement points and additional needs. Nonetheless, future work should extend TagDebt by implementing additional requirements from the original list and by refining the requirements through further empirical studies in diverse organizational settings.

There are two other threats related to the Design phase: the selection of the platform to run the bot and the selection of the NLP-based solutions. Regarding the selection of the platform, other platforms such as Jira\footnote{\url{https://www.atlassian.com/software/jira}} could have been chosen, and this selection could affect the perception of the bot by developers. However, GitHub is one of the largest software repositories, and most of the works related to TD identification are carried out using GitHub~\citep{Phaithoon2021, Yetistiren2022, Mohayeji2022}. Furthermore, GitHub has several bots and tools that developers use extensively. Therefore, we mitigate this threat by focusing on a well-known platform. As for the selection of the detection function, Li et al.'s model~\citep{Li2022} is a model specific for identifying SATD in issues. Although its accuracy is not too high (around 70\%), we deemed it fair for an initial version of our bot. To mitigate this threat, we developed a plugin-based system that, as already stated before, enables practitioners to easily replace the model, thereby increasing the potential for practitioners to adopt the bot. The plugin's flexibility is made explicit, since we also provide an LLM-based alternative, which implements GPT-5-mini. This LLM was selected because it is more affordable than the higher-hyperparameter counterparts, while providing good performance for TD classification, e.g., as recently reported by \cite{maarleveld26kubernetes}.

There is also a threat related to the focus on issues rather than other software artifacts (e.g., source code or pull requests). This decision might also impact the usefulness of the bot. To mitigate this threat, we based our bot on a scenario presented in our previous study~\citep{Biazotto2025b}, which was reviewed by practitioners who considered a bot in issue tracking systems as useful for TDM. 

The \textbf{Investigation} phase is key for DSR, as it provides empirical evidence on the developed artifact. In our context, we understand the perceptions of practitioners about the bot. To carry out our evaluation, we used a TAM-based method, and data analysis was carried out using thematic analysis, which poses its own set of threats. First, the study findings depend to some extent on how participants were selected and how they interpreted the questions in the interview. To mitigate this, we relied on a semi-structured interview protocol and encouraged open-ended discussions, allowing participants to express their critical views freely. In addition, each participant was interviewed only once. There is a threat related to the limited depth of insight from single-session interviews. Although no follow-up interviews were conducted, we addressed this limitation by thoroughly transcribing and coding the interviews to capture as much nuance as possible. Besides, participants were expected to use the bot for a few days before the interview. This allowed them to review the bot at their own pace and interact with it as much as they wanted.

Still regarding the \textbf{Investigation} phase, there is a threat related to the way we adapted TAM constructs into interview questions. Classical TAM instruments typically use Likert-scale items framed in a positive way (e.g., statements about whether a system is helpful or easy to use). In our qualitative setup, we kept this positive framing and adapted those items into closed questions (e.g., ``Do you think the bot would help to identify and monitor TD items more quickly and easily?'' or ``Was it easy to learn how to operate the bot?''). While these questions were always followed by open-ended and neutral probes (e.g., ``In which situations would it not help?''), the initial wording may still have encouraged participants to focus first on benefits. We mitigated this risk by systematically asking for drawbacks, conditions under which the bot would not be appropriate, and suggestions for improvement, and by incorporating both positive and negative indicators into our thematic analysis (e.g., concerns about precision, team size, configuration complexity, and documentation limitations). We deem that such mitigation actions were sufficient to reduce, at least to some extent, the potential bias caused by the set of questions.

We decided not to explicitly characterize participants’ working environments, and our results may be more difficult to interpret with respect to specific organizational or project conditions, and this may affect the transferability of the findings. In particular, perceptions of usefulness and ease of use can be shaped by factors such as organizational maturity, existing TD management practices, socio-technical constraints, and the project ecosystem; thus, the lack of structured contextual profiling may limit external validity. To mitigate this limitation, we asked participants to justify their answers and analyzed the contextual factors they reported as influencing adoption (e.g., team size, project characteristics, and codebase complexity). Future work should incorporate a more controlled in-project deployment and collect richer organizational and project-level descriptors to better explain acceptance across contexts.

To code the data in our study, we adopted an integrated approach~\citep{Cruzes2011}. Initially, two authors independently coded a randomly selected sample of 16 responses, generating as many codes as possible based on the data. They then discussed the generated codes until they reached a consensus. We followed a similar process for defining the themes and building the model. The first author proposed themes based on indicators from the data, which were subsequently discussed and reviewed by the other three authors. This approach allowed us to have more flexibility in code assignment and also to deepen our understanding of the data while discussing the codes and topics. In addition, we were concerned with defining themes that captured all aspects of the data. However, because we did not have categories to compare in the first round of coding (involving two authors), and because only one author coded the full interview set, we are not able to calculate an inter-rater agreement metric (e.g., Cohen's Kappa). Nonetheless, since all four authors thoroughly discussed the codes and themes, which evolved organically throughout the data analysis, we are confident that the lack of inter-rater agreement was sufficiently mitigated.

Another threat to the generalizability of our findings is that we relied on convenience sampling, inviting participants from our professional network. In this study, we relied on semi-structured interviews to enable in-depth discussions about TagDebt and its interactions with practitioners; however, this research method limited the number of participants we could recruit, which may reduce the diversity of perspectives and affect generalizability. To mitigate this threat, we followed the sampling and recruitment strategy described in Section~\ref{sec:sd}, recruiting participants with different roles (e.g., software engineers, architects, Q\&A analysts, and team leaders), with varying levels of experience (from 2 to 15 years), and with varied backgrounds and experience in TDM, ensuring that all interviewees met our inclusion criteria (i.e., experience with software development in industry). These decisions helped us capture a broad range of insights, even though the sample size and sampling strategy still pose limitations. Nonetheless, we acknowledge that broader longitudinal studies or large-scale replications are needed to further strengthen external validity. Furthermore, the bot was developed for GitHub Issues, which may limit the applicability of the findings to teams using other platforms (e.g., Jira, GitLab).

Regarding our methods for data analysis, although we used coding, which is a technique present in Grounded Theory~\citep{Strauss1990} to analyze the practitioners' answers and define the themes (as presented in Section~\ref{sec:evaluation-design}), not all steps of Grounded Theory were followed. Specifically, data collection and analysis steps were not concurrent (i.e., we collected all the data before analyzing it), which prevented us from claiming the theoretical saturation of our themes. Hence, replications of our study can help capture potential themes not identified in the current version of the study and contribute to advancing the understanding of TagDebt toward theoretical saturation. To support such replications and reproductions, we created a replication package\rp with all the necessary data and scripts to run the analyses.

Finally, it is also relevant to discuss \textbf{Reflexivity}. The authors have extensive research experience in TDM and were directly involved in the conception and development of TagDebt. While this expertise is valuable for framing the study and interpreting practitioners' feedback, it might also introduce a risk of confirmation bias (e.g., focusing on positive experiences or overlooking critical remarks). To mitigate this threat, we adopted several strategies. First, the data collection followed a semi-structured protocol, with open-ended follow-up questions that explicitly invited participants to report both benefits and drawbacks. Second, we used a bottom-up thematic analysis and grounded our codes and themes in verbatim quotations, preserving practitioners’ wording as much as possible. Third, all authors participated in the coding and theme discussions, reducing bias. While some residual influence of our positions cannot be fully excluded, these procedures were designed to make our interpretive process more transparent and to limit the impact of our prior expectations on the reported findings.

\section{Conclusion and Future Work}
\label{sec:conclusion}

A core lesson we learned while developing and evaluating TagDebt is that TDM tools must be easy to configure, easily integrated with existing workflows, and keep the human in the loop. According to our evaluation, TagDebt was shown to help improve the visibility of TD, as it increases the team's awareness about the TD items, leading to more informed decisions. The automation provided by TagDebt reduces repetitive manual tasks, enabling developers to focus on tasks that add more value to companies. In addition, TDM tools must focus on helping developers avoid errors and standardize issue reporting. 

Contextual factors play an important role in the intention of practitioners to use TagDebt, as well as other TDM tools. Larger teams and more complex projects benefit more from automation, whereas smaller teams often perceive less value. In addition, the roles of the practitioners, the type of project, and organizational privacy policies influence adoption decisions. These insights emphasize the importance of tailoring TDM tools to diverse development environments.

The contributions of this study open up several research directions for future work. First, expanding the bot’s analytical capabilities by integrating additional project artifacts (e.g., code, documentation) could improve classification precision and contextual relevance. Second, designing more intuitive interaction models, beyond comment-based commands, may enhance usability and encourage sustained use. Third, investigating the scalability and impact of such automation in large and complex teams can provide practical guidelines for organizational adoption.

As future work, we also plan to conduct a controlled comparative evaluation of TagDebt against alternative approaches for TD labeling and surfacing in GitHub, such as tag-based bots that detect explicit annotations (e.g., \#TODO/\#FIXME) and simpler labeling mechanisms supported by repository conventions. This study will require a curated set of projects and issues, controlled tasks, and consistent ground truth so that we can assess differences in user preference, perceived effort, workflow disruption, and correctness of produced labels. Such a comparative evaluation will complement our current practitioner-centered investigation by quantifying the benefits and trade-offs of TagDebt relative to existing baselines under comparable conditions.

In future work, we also plan to strengthen the empirical grounding of TagDebt by triangulating its evaluation with complementary tools and approaches that support other stages of TDM. For instance, we intend to evaluate TagDebt alongside static-analysis-based TD identification tools, or prioritization and repayment support approaches. Conducting such comparisons requires a different study design, including multiple artifacts, controlled tasks, and consistent project contexts to ensure fairness and interpretability, and therefore, it was outside the scope of our current exploratory, human-centric evaluation. By framing TagDebt as a step toward workflow-integrated TDM support, we aim to use these future comparative studies not only to quantify trade-offs across granularity levels (issue-level versus code-level) but also to guide concrete improvements to TagDebt, such as richer context-aware explanations and more actionable feedback.

Another potential research direction is a second evaluation round using UTAUT constructs with a more controlled and possibly longitudinal design. Such an evaluation could explicitly address constructs such as social influence and facilitating conditions when the goal shifts from exploratory feasibility to organizational-scale adoption. In addition, in such longitudinal studies, it is also possible to assess the real-world impact of automated TDM tools on software quality, team productivity, and decision-making to validate and refine their design. Such a study could employ TagDebt in one or more projects for an extended period, observing its usage and measuring whether and how it influences practitioners' decision-making. Besides, we can check whether TagDebt affects other TDM activities, such as prioritization and repayment. We can also investigate if the usage of TagDebt can reduce the accumulation of TD over time. Finally, the development of standardized taxonomies and customizable labeling frameworks could improve communication between heterogeneous teams and projects.

\section{Declarations}

\subsection{Funding}
This study was financed in part by the Brazilian Federal Agency for Support and Evaluation of Graduate Education (CAPES) - Finance Code 001, São Paulo Research Foundation (FAPESP) 2023/0488-5, and National Council for Scientific and Technological Development (CNPq) 313245/2021-5.

\subsection{Ethical approval}
Our study follows the core principles outlined in the ACM SIGSOFT empirical standard ``Ethics (Studies with Human Participants)'' (ACM SIGSOFT, 2024a). No risks of harm to participants were foreseen. The activity for evaluating our proposed tool required an estimated commitment of 30 minutes, comparable in scope and duration to programming assessments commonly applied in recruitment processes.
\\
\\
References:\\
ACM SIGSOFT (2024a) Ethics (Studies with Human Participants). \url{https://github.com/acmsigsoft/EmpiricalStandards/blob/master/docs/supplements/EthicsHumanParticipants.md} (Commit d131d720)

\subsection{Informed consent}
The participants were informed about the research project and received all necessary details to ensure their understanding before giving their consent. They were informed that participation is entirely voluntary and that they may withdraw from the study at any time without providing a reason, as long as this occurs before the first scientific output is submitted for publication.

It was explained that participation involves participating in an interview, during which data, such as role and experience, will be collected. Participants were informed about how data would be processed and that recordings would be deleted as soon as the transcripts were generated. 
In addition, participants were asked for specific consents:

\begin{itemize}
    \item Agreement for the anonymized transcript of the interview to be deposited in an open repository (such as Zenodo) so it can be reused for future research and learning purposes.
    \item Agreement to be contacted for future studies.
\end{itemize}

\subsection{Author Contributions}
All authors contributed to the study conception and design. Material preparation, data collection and analysis were performed by Jo\~ao Paulo Biazotto, and reviewed by the other three authors. The first draft of the manuscript was written by Jo\~ao Paulo Biazotto and all authors commented on previous versions of the manuscript. All authors read and approved the final manuscript.

\subsection{Data Availability Statement}
The generated datasets and scripts for data analysis are available in a \textit{Zenodo} repository at \href{https://doi.org/10.5281/zenodo.16934566}{https://doi.org/10.5281/zenodo.16934566}. Alternatively, the same replication package is also available on Github repository: \href{https://github.com/biazottoj/rp-tag-debt-bot}{https://github.com/biazottoj/rp-tag-debt-bot}

\subsection{Conflict of Interest}
The authors declared that they have no conflict of interest.

\subsection{Clinical Trial Number}
Clinical trial number: not applicable.

\bibliography{references}

\end{document}